\documentclass[12pt]{article} 
\pdfoutput=1
     \usepackage{amsmath}
  \usepackage{graphicx}
  \usepackage{color}
  \newlength{\abstractwidth}
 \usepackage{amsmath,amssymb}
\usepackage{graphicx}
\usepackage{subfigure}
\usepackage{caption}
\usepackage{lipsum}
\usepackage[utf8]{inputenc}
\usepackage{scalerel}
  \usepackage{mciteplus} 
\usepackage{hyperref}

\DeclareMathOperator{\Tr}{Tr}

  \newcommand{\be}{\begin{equation}}
  \newcommand{\bea}{\begin{eqnarray}}
  \newcommand{\eea}{\end{eqnarray}}
  \newcommand{\beq}{\begin{equation}}
  \newcommand{\ee}{\end{equation}}
  \newcommand{\eeq}{\end{equation}}

  \newcommand{\half}{{1\over 2}}
  
\def\la{\label}

  \def\ba{\begin{eqnarray}}
  \def\ea{\end{eqnarray}}

 \def\simleq{\; \raise0.3ex\hbox{$<$\kern-0.75em
      \raise-1.1ex\hbox{$\sim$}}\; }
 \def\simgeq{\; \raise0.3ex\hbox{$>$\kern-0.75em
      \raise-1.1ex\hbox{$\sim$}}\; }

\def\nref#1{(\ref{#1})}

\begin{document}

\begin{titlepage}
  \bigskip

  \bigskip\bigskip

  \bigskip

\begin{center}
{\Large \bf{The Quantum Gravity Dynamics of Near Extremal Black Holes}}
 \bigskip
{\Large \bf { }} 
    \bigskip
\bigskip
\end{center}

  \begin{center}

 \bf {Zhenbin Yang  }
  \bigskip \rm
\bigskip
 
 \rm

\bigskip
\bigskip
Jadwin Hall, Princeton University,  Princeton, NJ 08540, USA

  \end{center}

 \bigskip\bigskip
  \begin{abstract}

  We study the quantum effects of Near-Extremal black holes near their horizons. The gravitational dynamics in such backgrounds are closely connected to a particle in $AdS_2$ with constant electric field. We use this picture to solve the theory exactly. We will give a formula to calculate all correlation functions with quantum gravity backreactions as well as the exact Wheeler-DeWitt wavefunction.
  Using the WdW wavefunction, we investigate the complexity growth in quantum gravity.

 \medskip
  \noindent
  \end{abstract}
\bigskip \bigskip \bigskip

  \end{titlepage}

   \tableofcontents
\newpage
 \section{Introduction}
  Near-Extremal black holes have a  universal structure near their horizons: 
  there is an $AdS_2$ throat with a slowly varying internal space.  
 Its low energy  gravitational dynamics is captured universally by the following effective action in two dimensions \cite{Strominger:1994tn}:
 \begin{equation}\label{full action}
     I=\underbrace{-{\phi_0\over 2
     }\left(\int R+2\int_{\partial_M}K\right)}_{\text{Einstein-Hilbert Action}}\underbrace{-{1\over 2
     }\left(\int_M \phi(R+2)+2\int_{\partial M}\phi_b K\right)}_{\text{Jackiw-Teitelboim action}}+S_{matter}(g,\psi),
 \end{equation}
 where the dilaton field $\phi+\phi_0$ represents the size of internal space. We have separated the size of internal space into two parts: $\phi_0$ is its value at extremality. It sets the value of the extremal entropy
 which comes from the first term in \nref{full action}. 
 $\phi$ is the deviaton from this value. 
 We have also added matter that only couples to the metric. This is a reasonable assumption when matter comes 
 from Kaluza Klein reduction, where the coupling to the dilaton would involve $\phi/\phi_0 \ll 1$.

  The action $\int \phi(R+2)+2\int\phi_b K$ is the so-called Jackiw-Teitelboim action \cite{JACKIW1985343,TEITELBOIM198341}, and will be the main focus of our paper.
   This action is one of the simplest nontrivial gravitational actions in two dimensions.\footnote{Another nontrivial action is the CGHS model which could be written as $\int (\phi R+C)$, and that characterizes the horizon structure of general black holes.}.  It is simple because the bulk geometry is a rigid $AdS_2$ space fixed by the equation of motion of the dilaton field.  Its nontriviality arises from the remaining boundary action.  Schematically, the gravitational action is reduced to the following form:
 \begin{equation}\label{schematic action}
 I=-{ 2 \pi \phi_0 
 }\chi(M)-{\phi_b
 }\int_{\partial M}K +S_{matter}(g,\psi).     
 \end{equation}
 And the motion of the boundary is controlled by its extrinsic curvature.
  Our goal will be to quantize this action and to provide expression for the full quantum gravity correlators 
  of \nref{full action}. 
  This problem was considered before in \cite{Bagrets:2016cdf,Bagrets:2017pwq,Stanford:2017thb,Mertens:2017mtv,Kitaev:2018wpr} from various points of view. Here we will add one other point of view where we 
  reduce the problem to the motion of a relativistic particle in an electric field, building on a suggestion in 
  Kitaev's talk at IAS \cite{kitaevIAS}. More precisely, one can consider a relativistic particle in a Lorentzian $AdS_2$ target space moving under the influence of an electric field. The coupling to electric field can also be viewed as a coupling to a spin connection so that it becomes a particle with spin as suggested by Kitaev. Alternatively we can start from a non-relativistic particle moving in hyperbolic space, $H_2$, under the influence of a magnetic field $b$. After analytic 
  continuation in $b$ to imaginary values we get the problem of interest. 
  
  Using this point of view one can think of the full quantum gravity problem as the combination of two problems.
  First we consider quantum fields propagating in $AdS_2$ (or $H_2$ in the Euclidean case) and then we 
  add the ``gravitational particle" which couples to the quantum fields by changing their boundary location in $AdS_2$.
  The discussion of quantum fields will be standard and depends on the particular model one interested in, therefore we will mainly focus on solving the second problem.
  Generically, solving the gravitational problem is challenging and is not exactly equivalent to a quantum mechanical particle.  One needs to worry about what functional space one will integrate over.  For example, in path integrals, one usually integrates over all trajectories including those with self-intersections.  However self-intersecting boundaries in gravitational system have no obvious meaning.  On that account, more precisely the gravitational problem is equal to a self-avioding particle. 
  Nevertheless, it turns out that one can take a particular limit of this model, namely large $\phi_b$, to avoid this issue and a treatment of the boundary theory as an ordinary particle is justified.   It is also true that the JT gravity can be rewritten as a Schwarzian action only in this limit.  We call this the Schwarzian limit and will only focus on solving the JT action in the Schwarzian limit. Solving the model away from Schwarzian limit was considered recently by Kitaev and Suh \cite{Kitaev:2018wpr}.
  
  Our result can be summarized as follows: 
  
  \textsl{First, we will give a formula to calculate all correlation functions with quantum gravity backreaction (formula \ref{final formula}). Second, we will give the exact Wheeler-DeWitt wavefunction in the Schwarzian limit, which has been analyzed classically by Harlow and Jafferis \cite{Harlow:2018tqv}. Last, we consider the recent proposed conjecture about complexity growth in this exact Wheeler-DeWitt wavefunction and show that the complexity maintains linear growth after taking quantum gravity effects into account.  This, to our knowledge, is the first test of the gravitational conjecture made by Susskind \cite{Susskind:2018fmx} that the size of ERB grows linearly for as long as quantum mechanics allows.}
  
 This paper is organized as follows: in the first section, we will review the classical calculation of this model and introduce notations; in the second section, we will make the dictionary between the JT model and a particle in a magnetic field; in the third and fourth sections, we will solve the quantum mechanical problem and derive the propagator and WdW wavefunction in the Schwarzian limit; in the last section, we will talk about gravitaional backreaction on correlators as well as complexity growth.  As a useful notion in our calculation, we introduce a notation called gravitational Feyman diagrams. 
  
 \section{Classical Solutions}
  Let us first consider the classical solutions of 
  the Jackiw-Teitelboim model \nref{full action}. See
  \cite{Almheiri:2014cka} for further discussion. 
 The equation of motion of the dilaton field imposes $R=-2$
 and 
 fixes the geometry to be $AdS_2$, or $H_2$ in the Euclidean case.
 This is also true if
   we have additional matter coupled with metric only, as in
   \nref{full action}.
 The   equations for the metric constrain the dilaton 
  \begin{equation}\label{Einstein Equation}
     (\nabla_{\mu}\nabla_{\nu}\phi-g_{\mu\nu}\nabla^2\phi+g_{\mu\nu}\phi)+T^{M}_{\mu\nu}=0
 \end{equation}
 These equations are compatible with each other thanks to the conservation of the matter stress tensor. They do not allow
 any propagating mode. In fact,  setting $T^M_{\mu\nu}=0$, and using a high momentum approximation we can write 
 \nref{Einstein Equation} as $ (k_{\mu}k_{\nu}-g_{\mu\nu}k^2)\phi(\vec k)=0 $, which then implies   $  \phi(\vec k)=0 $, for large $k$.

More precisely, after introducing the Euclidean $AdS_2$ coordinates $ds^2 = d\rho^2 + \sinh^2\rho d\varphi^2$, we can solve 
  (\ref{Einstein Equation}) in $AdS_2$ with no matter. Up to an SL(2) transformation the solution is 
\begin{equation}
    \phi=\phi_h \cosh \rho,
\end{equation}
where $\phi_h$ is a constant that is fixed by the boundary conditions. 
At the boundary we fix the metric along the boundary and the value of the dilaton field 
\be \la{BCond}
ds_{\parallel} = d u { \phi_r \over \epsilon }   ~,~~~~~~~~~~~ \phi= \phi_b = { \phi_r \over \epsilon }
\ee
where we think of $u$ as the time of the boundary theory. It is simply a rescaled version of proper time. Similarly $\phi_r$ is a rescaled
value of the dilaton. We will be interested in taking $\epsilon \to 0$. With these rescalings the value of $\phi_h$ in the interior 
remains fixed as we take $\epsilon \to 0$ as $\phi_h =  2 \pi /\beta $ where $\beta$ is the inverse temperature, $u \sim u + \beta$. Notice that due to the factor of $\phi_r$ in the first expression in \nref{BCond} we are measuring time in units of the  constant $\phi_r$, which has dimensions of length. We did this for convenience. 
A nice feature that appears after taking the $\epsilon \to 0$ limit is that the action (\ref{schematic action}) can be written as the Schwarzian action for the boundary curve labeled by $\varphi(u)$ \cite{Maldacena:2016upp}:
\begin{equation}
I=-\int du Sch(\tan{\varphi(u)\over 2},u)
\end{equation}
 The fluctuation of the boundary shape can be understood as the fluctuation of the dilaton distribution in the bulk.
A bit more explicitly we can say that the dilaton boundary condition fixes the location of the boundary at $\rho_b$ given by 
$\phi_b = \phi_h \cosh \rho_b$, and the metric at that location relates the time $\varphi$ to $u$ by $\phi_r du = \epsilon \sinh \rho_b d\varphi $.
We get the above formulas noticing that the period of $\varphi$ is $2\pi$ while that of $u$ is $\beta$, which fixes $\epsilon \sinh \rho_b$.

\section{Charged Particle in $AdS_2$}

Despite the absence of a bulk propagating mode there is still a non-trivial dynamical gravitational degree of freedom. There are various 
ways to describe it. Here we will think of it as arising from the motion of the physical boundary of $AdS_2$ inside a rigid $AdS_2$ space. 
This picture is most clear for finite $\epsilon$ in \nref{BCond}, but it is true even as $\epsilon \to 0$. The dynamics of the boundary 
is SL(2) invariant. This SL(2) invariance is a gauge symmetry since it simply reflects the freedom we have for cutting out a piece
of $AdS_2$ space that we will call the ``inside". It is important that the dilaton field we discussed above is produced after we put in the 
boundary and it moves together with the boundary under this SL(2) gauge transformation. It is a bit like the Mach principle, the location
in $AdS_2$ is only defined after we fix the boundary (or distant ``stars"). 

We can make this picture of a dynamical boundary more manifest as follows. 
 Since the bulk  Jackiw-Teitelboim action \nref{full action} is linear in $\phi$, we can integrate out the dilaton field which 
 sets the metric to that of $AdS_2$ and removes the bulk term in the action, leaving only the term involving the extrinsic curvature
  \bea
  I&=&-{\phi_r\over
  \epsilon}\int du \sqrt{g}K
  \eea
  This action, however, is divergent as we take $\epsilon$ to zero. This divergence is simply proportional to the length of the boundary 
  and can be interpreted as a contribution to the ground state energy of the system. 
 So we introduce a counterterm proportional to the length of the boundary to cancel it. This is just a common shift of the energies 
 of all states. 
 It is also convenient to use the 
 Gauss-Bonnet theorem to relate the extrinsic curvature to an integral over the bulk 
  \begin{equation}
      \int_{\partial_M} du \sqrt{g}K=2\pi\chi(M)-{1\over 2}\int_M R
  \end{equation}
  Since the curvature is a constant, the bulk integral is actually proportional to the total area $A$ of our space.  That is we have the regularized action:
  \bea\label{subtraction}
  I&=&-{\phi_r\over 
  \epsilon}\int_{\partial M} du \sqrt{g}(K-\underbrace{1}_{\text{counterterm}})=-{\phi_r\over
  \epsilon}\left(2 \pi\chi(M)-{1\over 2}\int_{M}R-\int_{\partial M} du \sqrt{g}
  \right) \nonumber\\
 &=&-2\pi q\chi(M)-q A+q L,~~~~~~~~~~~~~~~~~~~~~~~~~~~q\equiv{\phi_r\over 
 \epsilon} ,~~L={\beta \phi_r\over \epsilon}
  \eea
 We now define an external gauge field $a_\mu$ as 
\begin{equation}
a_{\varphi}=\cosh\rho-1,~~~~~~~a_{\rho}=0,~~~~~~~f_{\rho \varphi}=\sinh\rho=\sqrt{g},
\end{equation}  
and write the action as follows
\begin{equation}\label{regularized action}
    I= - 2 \pi q + q L - q \int a
\end{equation}
where we used that  $ \chi(M)$ is a topological invariant equal to one, in our case, where the topology is that of a disk. 
The term $q L$ is just the length of the boundary. So this action has a form somewhat 
similar to the action of a relativistic charged particle moving in 
$AdS_2$ in the presence of a constant electric field.
There are a couple of  important differences. First we are summing only over trajectories of fixed proper length set by the inverse temperature 
$\beta$. Second, in the JT theory we are treating the $SL(2)$ symmetry as a gauge symmetry. And finally, in the JT theory we identify the proper length with the boundary time, viewing configurations which differ only by a shift in proper time as inequivalent. In fact, all these changes simplify the problem: we can actually think 
of the problem as a non-relativistic particle moving on $H_2$ in an electric field. In appendix
\ref{appendix:relativistic particle} we discuss in more detail the connection to the relativistic particle. 

In fact, precisely the problem we are interested in has been discussed by Polyakov in \cite{Polyakov:1987ez}, Chapter 9, as an an intermediate step for the sum over paths. 
Now we would also like to point out that we can directly get to the final formula by 
using the discussion there, where he explicitly shows
that for a particle in flat space the sum over paths of fixed proper length that stretch between 
two points $\vec x$ and $\vec x'$ gives 
\be \la{PolFlat}
\int {\cal D }\vec x e^{ - m_0 \tilde\tau } \delta(  \dot { \vec x}^2  - 1    ) = 
e^{ - \half \mu^2  \tau}  \langle x' | e^{ - \tau H } |x\rangle = 
e^{ - \half \mu^2  \tau}  \int { \cal D}{\bf x} \exp\left( - \int_0^\tau d\tau' \half {  \dot \vec x^2   }  
\right)
\ee
 $\mu^2$ is the regularized mass and $\tilde \tau$ is related to $\tau$ by a multiplicative renormalization. 
The JT model consists precisely of a functional integral of this form, where we fix the proper length along the boundary. There are two simple modifications, first the particle is in a curved $H_2$ space and second we have the coupling to the electric field. These are minor modifications, but the arguments leading
to \nref{PolFlat} continue to be valid so that the partition function of the JT model can be
written directly:
\be \la{PolAdS}
\scaleto{\int {\cal D }\vec x e^{2\pi q - m_0 \tilde\tau +q\int a } \delta(  { \dot x^2 + \dot y^2 \over y^2 }- q^2     ) = 
e^{ 2\pi q- \half \mu^2  \tau} \Tr  e^{ - \tau H } = e^{2\pi q - \half \mu^2 \tau }
\int { \cal D}{\bf x} \exp\left( - \int_0^\tau d\tau' \half { \dot x^2 + \dot y^2 \over y^2 } - q {\dot x \over y}
\right)}{26pt}
\ee
 The delta function implements the first condition 
in \nref{BCond} at each point along the path. 
The last path integral can be done exactly by doing canonical quantization of the action (section \ref{CanonicalQuantization}) and by comparing the result with the one from the Schwarzian action \cite{Stanford:2017thb} 
we can determine that $\tau$ is the inverse temperature $\beta$.

In the above discussion we have been fixing the time along the boundary. Instead we can fix the energy 
at the boundary, where the energy is the variable conjugate to time. 
This can be done by simply integrating \nref{PolAdS} times $e^{ \beta E}$
over $\beta$ along the imaginary axis. This fixes the energy of the non-relativistic problem by generating a $\delta( H - E)$. More precisely, we will argue that after doing a spectral decomposition we can write
the propagator at coincident points as 
\be \la{JTEn}
Z_{JT}(\beta ) = \int_0^\infty \rho(E)e^{-\beta E} dE  \longrightarrow \rho(E) = \int_{-i \infty}^{i \infty}
{ d\beta \over i } e^{ E \beta } Z_{JT}(\beta) 
\ee
where the function $\rho(E) $ can then be interpreted as a ``density of states" in the microcannonical ensemble. We will give its explicit form in section (4.2). 
For now, we only want to contrast this integral with a superficially similar one that appears when we 
compute the relativistic propagator 
\be \la{RelEl}
 e^{-2\pi q} \int_0^\infty e^{ E \beta }Z_{JT} (\beta) = \langle \phi(x) \phi(x) \rangle 
 \ee
 which gives the relativistic propagator of a massive particle in an electric field at coincident points
 (we can also compute this at non-coincident points to get a finite answer). The total mass of the 
 particle is 
 \be
 m = q - {E\over q}    
 \ee
 For large $q$ this is above threshold for pair creation\footnote{See Appendix \ref{appendix:relativistic particle}}. The pair creation interpretation 
 is appropriate for the problem in \nref{RelEl}, but not for \nref{JTEn}. 
 In both problems we have a classical approximation to the dynamics that corresponds to a particle 
 describing a big circular trajectory in hyperbolic space at radius $\rho_c$ :
 \be \la{SolCir}
\tanh \rho_c = { m \over q }
  \ee
  For the problem in \nref{RelEl}, fluctuations around this circle lead to an instability, with a single negative mode and an imaginary part in the partition function \nref{RelEl}. This single negative mode corresponds to small fluctuations of the overall size of the circular trajectory around \nref{SolCir}.
  On the other hand in \nref{JTEn} we are integrating the same mode along a different contour, along the 
  imaginary axis, where we get a real and finite answer. Furthermore, the imaginary part in the 
  partition function \nref{RelEl} comes precisely from the trajectory describing pair creation, which is also the type of contribution captured in \nref{JTEn}.
  
  Finally, in the relativistic particle problem, we expect that the pair creation amplitude should be exponentially suppressed for large $q$, while the partition function for the JT model is not.  In fact, 
  for large $q$ the exponential suppression factor for pair creation goes as $e^{ - 2 \pi q } $, which is precisely cancelled by a similar factor in \nref{JTPa}, to obtain something finite in the large $q$ limit.

\section{Solving the Quantum Mechanical Problem}

\la{CanonicalQuantization}

As we explained above the solution of the JT theory is equivalent to considering a non-relativistic particle in $AdS_2$ or $H_2$. 
We first consider the Euclidean problem, of a particle moving in $H_2$. An ordinary magnetic field in $H_2$ leads to an Euclidean action  
of the form 
\be \la{magnetic field} 
S= \int du \half  { \dot x^2 + \dot y^2 \over y^2 } + i b \int d u  { \dot x \over y }   -\half  ( b^2 +{ 1 \over 4} ) \int d u ~,~~~~~~~b = i q
\ee
If $b$ is real we will call it a magnetic field, 
when $q$ is real we will call it an ``electric" field. The last term is a constant we introduced for convenience. Its only effect
will be to shift the ground state energy.  It is interesting to compute the classical solutions and the corresponding action for
\nref{magnetic field}. These solutions are simplest in the $\rho$ and $\varphi $ coordinates, using the SL(2) symmetry we find that the trajectories are given by ($t=-iu$):
\bea
{1\over 2}\sinh^2\rho ({d\varphi\over dt})^2+{q^2\over 2}-{1\over 8}=E,~~ \cosh\rho={q\beta\over 2\pi},~~ {d\varphi\over du}={2\pi\over \beta}.
\eea
In this classical limit we 
get the following relations for the action and the temperature:
\bea 
 & ~& { \beta \over 2 \pi } = {1\over \sqrt{2E+{1\over 4}}}
\cr
& ~ & -S = {2\pi^2\over \beta}+{\beta\over 8}-2\pi q \la{ClassEnt} 
\eea
        
When $b$ is real, this system is fairly conventional and it was solved in \cite{Comtet:1986ki} . Its detailed spectrum depends on $b$. For very large $b$ we have a series of Landau levels and also a continuous spectrum. In fact, already the classical problem contains closed circular orbits, related to the discrete Landau levels, as well as orbits that go all the way to infinity.\footnote{See Appendix \ref{appendix:landau level}}
The number of discrete Landau levels decreases as we decrease the magnetic field and for $0 < b < \half$ we only get a continuous spectrum.  The system has a $SL(2)$ symmetry and the spectrum organizes into SL(2) representations, which are all in the
continuous series for $0 < b < 1/2$. For real $q$ we also find a continuous spectrum which we can view as the analytic continuation 
of the one for this last range of $b$.

  The canonical momenta of the action (\ref{magnetic field}) are:
  \be
  p_x={\dot x\over y^2}+{i q\over y};~~~~~~~p_y={\dot y\over y^2}.
  \ee
  And the Hamiltonian conjugate to $\tau_L$ is thus:
  \be
  H={\dot x^2+\dot y^2\over 2y^2}+{q^2\over 2}={y^2\over 2} [(p_x-i{q\over y})^2+p_y^2] +{ q^2 \over 2 } - { 1 \over 8} 
  \ee
  Note that the Hamiltonian is not Hermitian. However, it is {\cal PT}-symmetric (here parity reflects $x$ and $p_x$) and for that
  reason the spectrum is still real, see \cite{Bender:2005tb}. 
  The action is invariant under $SL(2,R)$ transformations generated by
 \bea
 L_0=x p_x+y p_y;~~~~~L_{-1}=p_x;~~~~~~~~L_{1}=(y^2-x^2)p_x-2xyp_y-2iqy  \la{SLtwoGen}
 \eea
 Notice the extra $q$ dependent term in $L_1$ that arises due to the presence of a magnetic field. 
Up to a simple additive constant, the Hamiltonian is proportional to  the Casimir operator
 \be
 H={1\over 2}\left( L_0^2+ { 1 \over 2} L_{-1}L_{1} +  { 1 \over 2} L_{1}L_{-1}  \right)   + { q^2 \over 2} - { 1 \over 8 } 
 \ee
As is common practice,  let us label the states by quantum numbers $j={1\over 2}+is$ and $k$, so that $H|j,k\rangle=j(1-j)|j,k\rangle$ and $L_{-1}|j,k\rangle=k|j,k\rangle$. 
We can find the eigenfunctions by 
 solving the $Schr\ddot{o}dinger$ equation with boundary condition that the wavefunction should vanish at the horizon $y\rightarrow \infty$ \cite{Comtet:1984mm,Comtet:1986ki,Pioline:2005pf}:
  \be
  \omega_{s,k}={s^2\over 2} ,~~~~~~f_{s,k}(x,y)=
  \begin{cases}
({s\sinh 2\pi s\over 4\pi^3k})^{1\over 2}|\Gamma(is-b+{1\over 2})|e^{-ik x}W_{b,is}(2 k y), ~~~k>0;\\
({s\sinh 2\pi s\over 4\pi^3|k|})^{1\over 2}|\Gamma(is+b+{1\over 2})|e^{-ik x}W_{-b,is}(2|k|y),~~~k<0.
\end{cases}
  \ee
  where $\omega_{ks}$ is giving the energy of the states labelled by $s$ and $k$, and $ W $ is the Whittaker function. The additive constant in \nref{magnetic field} was
  introduced to simplify this equation. We can think of $s$ as the quantum number of the continuous series representation of $SL(2)$ with 
  spin $j=\half + i s $. 

After continuing $b \to i q$ we find that the gravitational system has a continuous 
spectrum 
\begin{equation}\label{energy relation}
    E(s)={s^2 \over 2 }.
\end{equation}

\subsection{The Propagator }

It is useful to compute the propagator for the non-relativistic particle in a magnetic field, 
$K(u,\boldsymbol{x_1},\boldsymbol{ x_2})=\langle \boldsymbol{x_1}| e^{-u H}|\boldsymbol{x_2}\rangle$. Here, $\boldsymbol{x}$ stands for ${x,y}$. 
The propagator for a real magnetic particle was obtained in \cite{Comtet:1986ki}:
\bea \label{particle heat kernel}
G(u,\boldsymbol{x_1},\boldsymbol{x_2})& =& e^{i \varphi(\boldsymbol{x_1},\boldsymbol{x_2})} \int_0^{\infty}ds s e^{- u { s^2 \over 2}}{\sinh 2\pi s\over 2\pi (\cosh 2\pi s+\cos 2\pi b)}{1\over d^{1+2is}}
\times 
\cr
& & ~~~~~ \times  ~_2F_1({1\over 2}-b+is,{1\over 2}+b+is,1,1-{1\over d^2}).
\cr
 d  & = & \sqrt{(x_1-x_2)^2+(y_1+y_2)^2\over 4y_1 y_2}
\cr
e^{ i \varphi(\boldsymbol{x_1},\boldsymbol{x_2})} &=& e^{-2 i b  \arctan {x_1-x_2\over y_1+y_2}}
\eea
In the case that we have a real magnetic field the prefactor is a phase and it is 
gauge dependent. It is equal to the value of Wilson Line $e^{i\int a}$ stretched along the geodesic between $x_2$ and $x_1$.
Here we quoted the value in the gauge where the action is 
\nref{magnetic field}. The second equation defines the parameter $d$, which is a function of the geodesic distance between the two points. Note that $d=1$ corresponds to coincident points.  
We can get the answer we want by making the analytic continuation $b\to i q$ of this formula. 
We can check that this is the right answer for our problem by noticing the following. First one can check that this expression is invariant under the SL(2) symmetry $L_a = L^1_a + L^2_a$ where $L_a$ are the generators \nref{SLtwoGen} acting on $\boldsymbol{x_1}$
and $L_a^2$ are similar generators as in \nref{SLtwoGen}, but with $q\to -q$. It is possible to commute 
the phase $e^{ i \varphi(\boldsymbol{x_1},\boldsymbol{x_2})}$,  in \nref{particle heat kernel} past these generators which would remove the $q$ dependent terms. This implies
that the rest should be a function of the proper distance, which is the case with \nref{particle heat kernel}. 
Then we can check the equation 
\bea
0 &=& (\partial_u  + H_1 )G(u,\boldsymbol{x_1},\boldsymbol{x_2}) 
\eea
which is also indeed obeyed by this expression. The $s$ dependent prefactor is fixed by the requirement that the 
propagator composes properly, or more precisely, by saying that for $u=0$ we should get a $\delta$ function. 

\subsection{Partition Function}

The gravitational partition function is related with the particle partition function with inverse temperature $\beta$.
\begin{figure}[h!]
\centering
\includegraphics[scale=0.3]{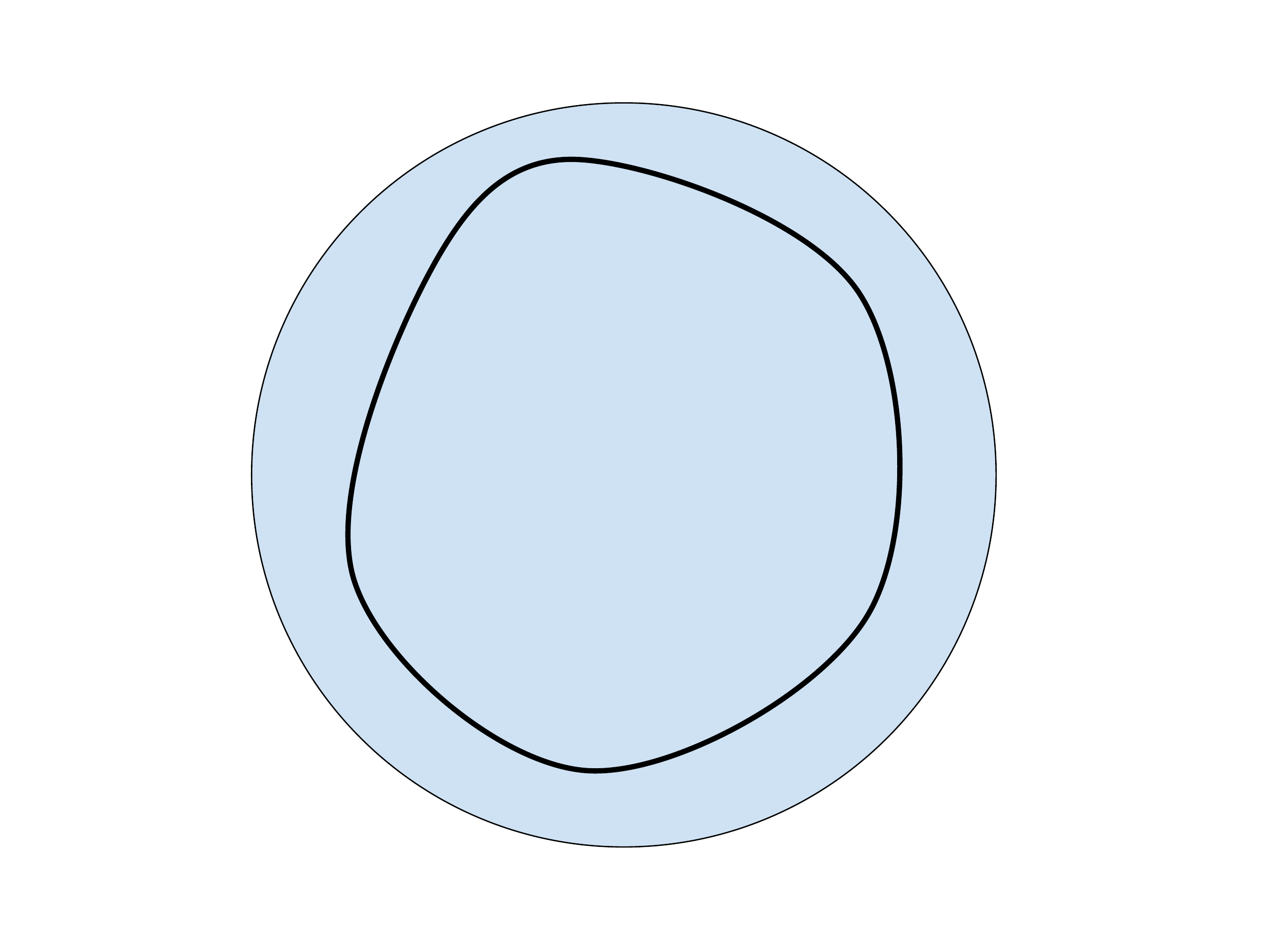}
\caption{Free Energy diagram with inverse temperature $\beta$.  }
\end{figure}
The canonical partition function of the quantum mechanical system is
  \bea
  Z_{\scaleto{Particle}{4pt}}&=&Tr e^{-\beta H}
  =\int_0^{\infty} ds \int_{-\infty}^{\infty}dk \int_M {d x d y\over y^2}  e^{-{\beta } { s^2 \over 2 }   }f^*_{s,k}(x,y)f_{s,k}(x,y)\nonumber\\
  &=&V_{AdS}\int_0^{\infty}ds e^{-\beta {s^2\over 2}}{s\over 2\pi}{\sinh(2\pi s)\over \cosh(2\pi q)+\cosh(2\pi s)}.
  \eea
The volume factor $V_{AdS}$ arises because after momentum integration there is no position dependence.  
In a normal quantum mechanical system, the volume factor means that the particle can have independent configurations at different locations of our space, however for a gravitational system this should be thought as redundant and should be cancelled by the volume of $SL(2,R) $ gauge group $2\pi V_{AdS}$\footnote{There might be a multiplicative factor in the volume of gauge group, but we can always absorb that into $S_0$.}.  
In gravitational system, there can also other contributions to the entropy from pure topological action. 
These give a contribution to the ground state entropy $S_0$. Including the topological action in (\ref{regularized action}), we find a  gravitational ``density of states" as 
\be \la{Spectral}
\rho(s)=\underbrace{e^{S_0}e^{2\pi q}}_{\text{extra terms}}\underbrace{1\over {2\pi}}_{\text{residue gauge}}\underbrace{{s\over 2\pi}{\sinh(2\pi s)\over \cosh(2\pi q)+\cosh(2\pi s)}}_{\text{particle in magnetic field}}=e^{S_0}e^{2\pi q}{s\over 2\pi^2}\sum_{k=1}^{\infty}(-1)^{k-1}e^{-2\pi q k}\sinh(2\pi s k).
\ee
 We have not given an explicit description of these states in the Lorentzian theory. More details were discussed in \cite{Kitaev:2018wpr,Lin:2018xkj}. 
 
 This expression has some interesting features. Notice that the classical limit corresponds to large $q$ and large $s$, where we reproduce \nref{ClassEnt}. After approximating, the density of states are 
 $\log \rho(s) \sim  S_0 + 2 \pi s $ for $ s/q < 1$ and $S_0+2\pi q$ for $s/q>1$.
 
 We can also expand the partition function for very small and very large temperatures where we obtain 
 \bea
 Z_{JT} &\sim & e^{ S_0} e^{ 2 \pi q } { 1 \over 4\pi^2\beta}  ~,~~~~~~~~~~~ \beta \ll {1\over q} 
 \cr
 Z_{JT} & \sim & e^{S_0} { 1 \over \sqrt{2\pi}\beta^{3/2} }~,~~~~~~~~~~~~ \beta \gg 1 
 \eea
 Notice that at leading order we get an almost constant entropy both at low and high temperatures, with the 
 high temperature one being higher. In both cases there are power law corrections in temperature.

\begin{figure*}[h!]
  \centering
  \subfigure[Density of States]{%
    \includegraphics[width=0.4\textwidth]{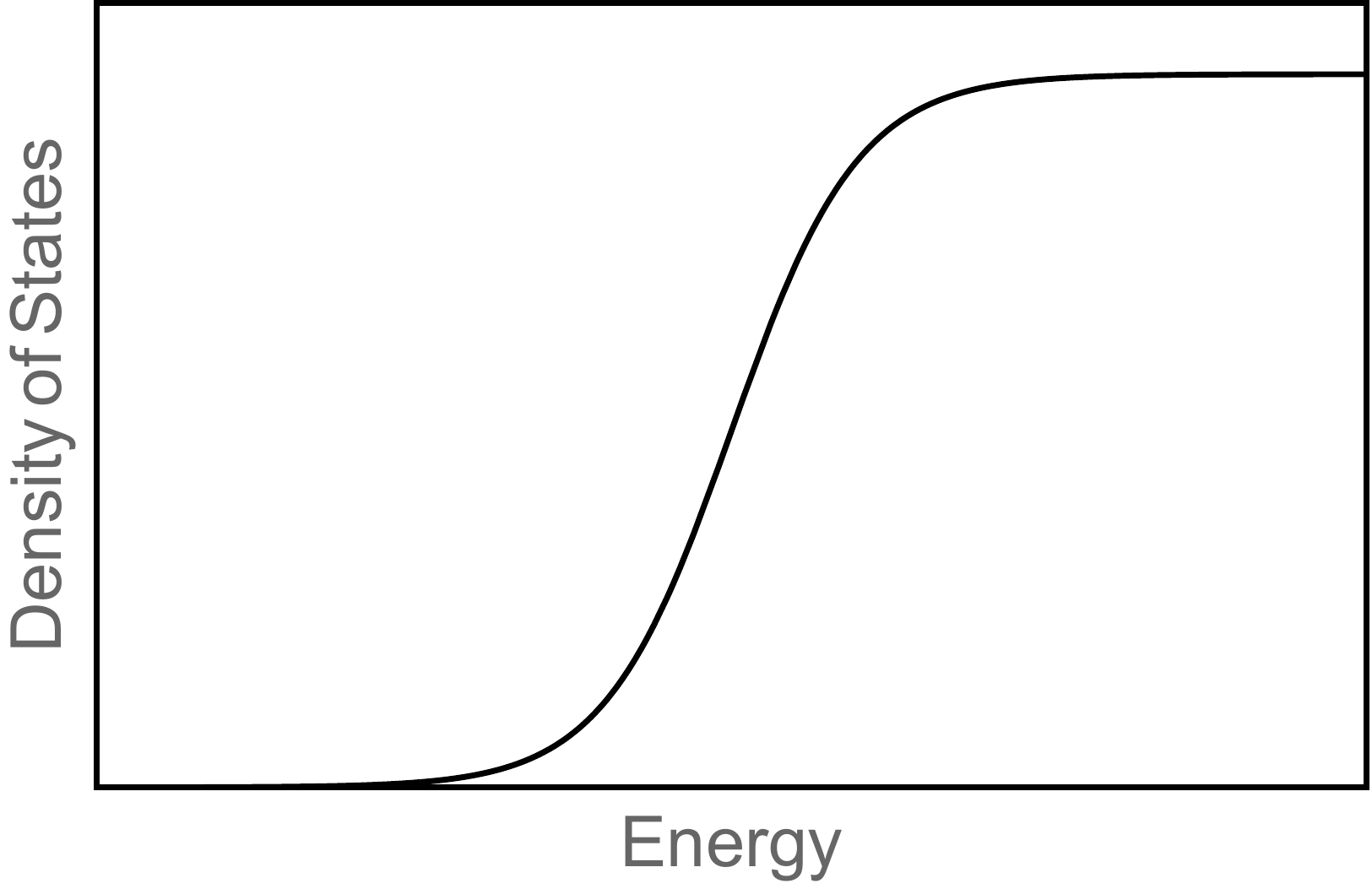}%
    \label{fig:dos}%
    }\hspace{4cm}
    \subfigure[Two Instantons]{%
    \includegraphics[width=0.35\textwidth]{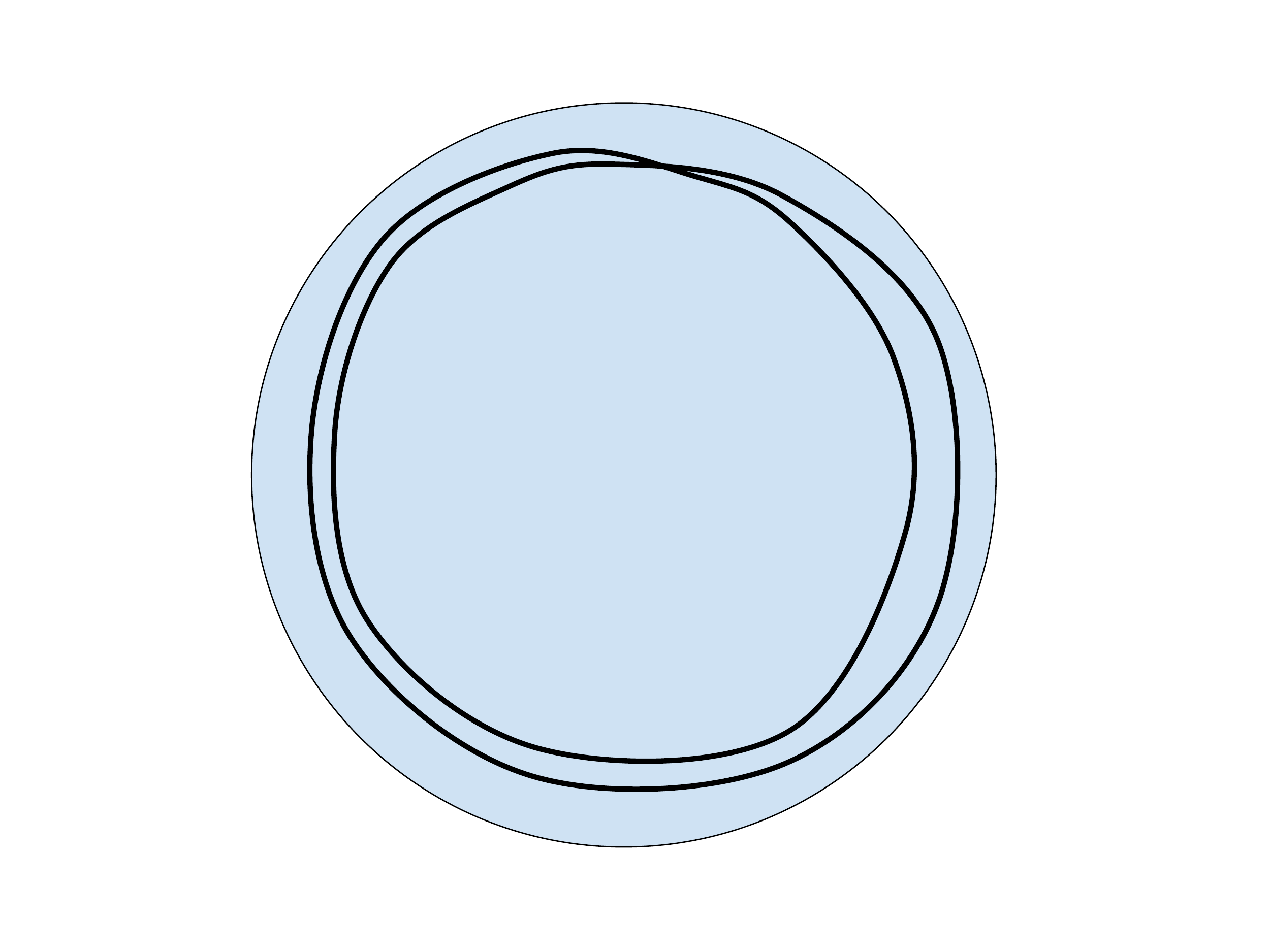}%
    \label{fig:TwoInstanton}%
  }%
  \caption{Density of States and the Two Instantons configuration} 

\end{figure*}

 Before we try to further elucidate the interpretation of this result, let us emphasize  a couple of important defects of our discussion. First, when we replaced the partition function of the JT theory by the action of a non-relativistic
 particle in an electric field, we were summing over paths in $H_2$. This includes paths that self intersect see figure \ref{fig:TwoInstanton}. Such paths do not have an obvious interpretation in the JT theory and it is not even clear that we should include them.  For example, the sum over $k$ in \nref{Spectral} can be understood in terms of classical solutions which wind $k$ times around the circle. These make sense for the problem of the particle in the electric field but apparently not in the JT theory.
 Maybe such paths could be given some interpretation in the gravity theory.
  Alternatively, we might want to sum over paths that do not self intersect. 
 A second defect is that we would be eventually interested in adding some matter fields propagating in the bulk geometry. These matter fields have boundary conditions at the boundary of the region of hyperbolic space 
 cut out by the boundary trajectory. The partition function of the fields with such an arbitrary boundary trajectory could also modify the results we described above. Of course, this issue does not arise if we have
 the pure JT theory. It is only important if we want to introduce bulk matter fields to define more complex observables.

 Instead of attempting to address the above issues, we will take an easy route, which is to consider the system only in the large $q$ (or small $\epsilon$) limit. In this regime, we address the above issues, and we can still trust the description of the particle in the electric field. 
 This large $q$ or small $\epsilon$ limit is the same one that isolates the Schwarzian action from the JT theory
 \cite{Maldacena:2016upp,Jensen:2016pah,Engelsoy:2016xyb}. 
 It turns out that the limit can be taken already at the level of the mechanical system, a simple rescaled
 version of the above system. This provides an alternative method for quantizing the Schwarzian theory. It has the advantage of being a straightforward second order action of a particle moving in a region near the boundary of hyperbolic. Of course, the Schwarzian theory was already quantized using a variety of methods in 
 \cite{Bagrets:2016cdf,Bagrets:2017pwq,Stanford:2017thb,Mertens:2017mtv,Luca}. We will simply provide yet another perspective, recover the old results, and write a few new expressions. 
  
\section{Quantum Gravity at Schwarzian Limit}
 
 Before getting into the details notice that the large $q$ limit of \nref{Spectral} gives 
 \be
 \rho(s) = e^{S_0}{s\over 2\pi^2}\sinh(2\pi s), ~~~E = { s^2 \over 2} ~,~~~~~~~Z_{JT} = \int_0^\infty ds 
 \rho(s) e^{ - \beta { s^2 \over 2 } } =e^{S_0}{1 \over\sqrt{2\pi}\beta^{3\over 2}}e^{2\pi^2  \over \beta}.
 \ee
 This reproduces what was found in \cite{Stanford:2017thb, Mertens:2017mtv, Mertens:2018fds} by 
 other methods. We see that we get a finite answer and also that the contributions from the $k>1$ terms in 
 \nref{Spectral} have disappeared. Because the $S_0$ part decouples with JT gravity, from now on, we will drop it and discuss $S_0$ only when it is necessary.

\subsection{The Propagator}

\begin{figure*}[h!]
  \centering
  \subfigure[singular ``phase" factor for different ordering]{%
    \includegraphics[width=0.45\textwidth]{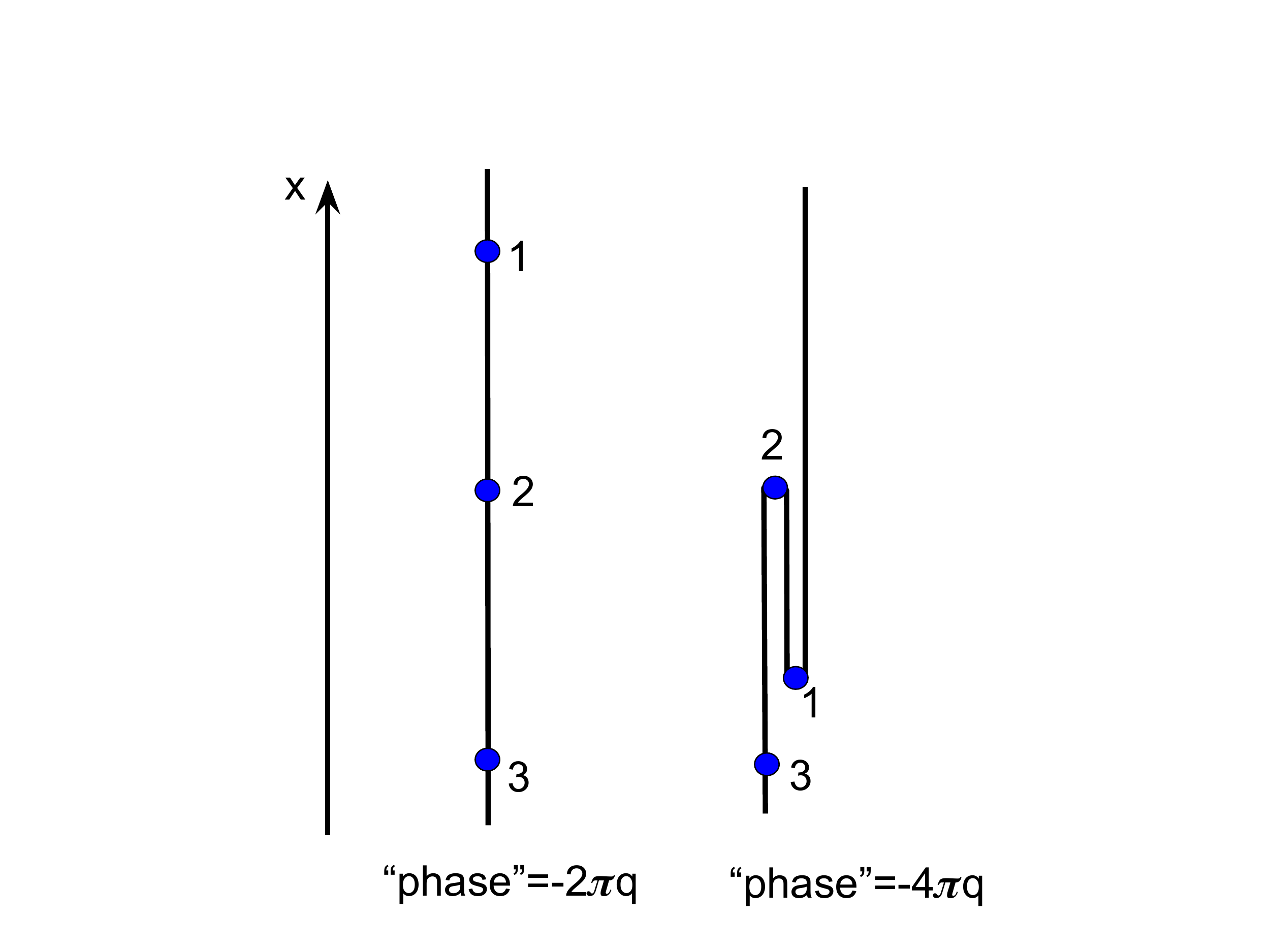}%
    \label{fig:phasefactor}%
    }\hspace{1.5cm}
    \subfigure[A geometric representation of the propagator. Here we fix the span angle $\theta$, the propagator is a summation over such geometries.]{%
    \includegraphics[width=0.45\textwidth]{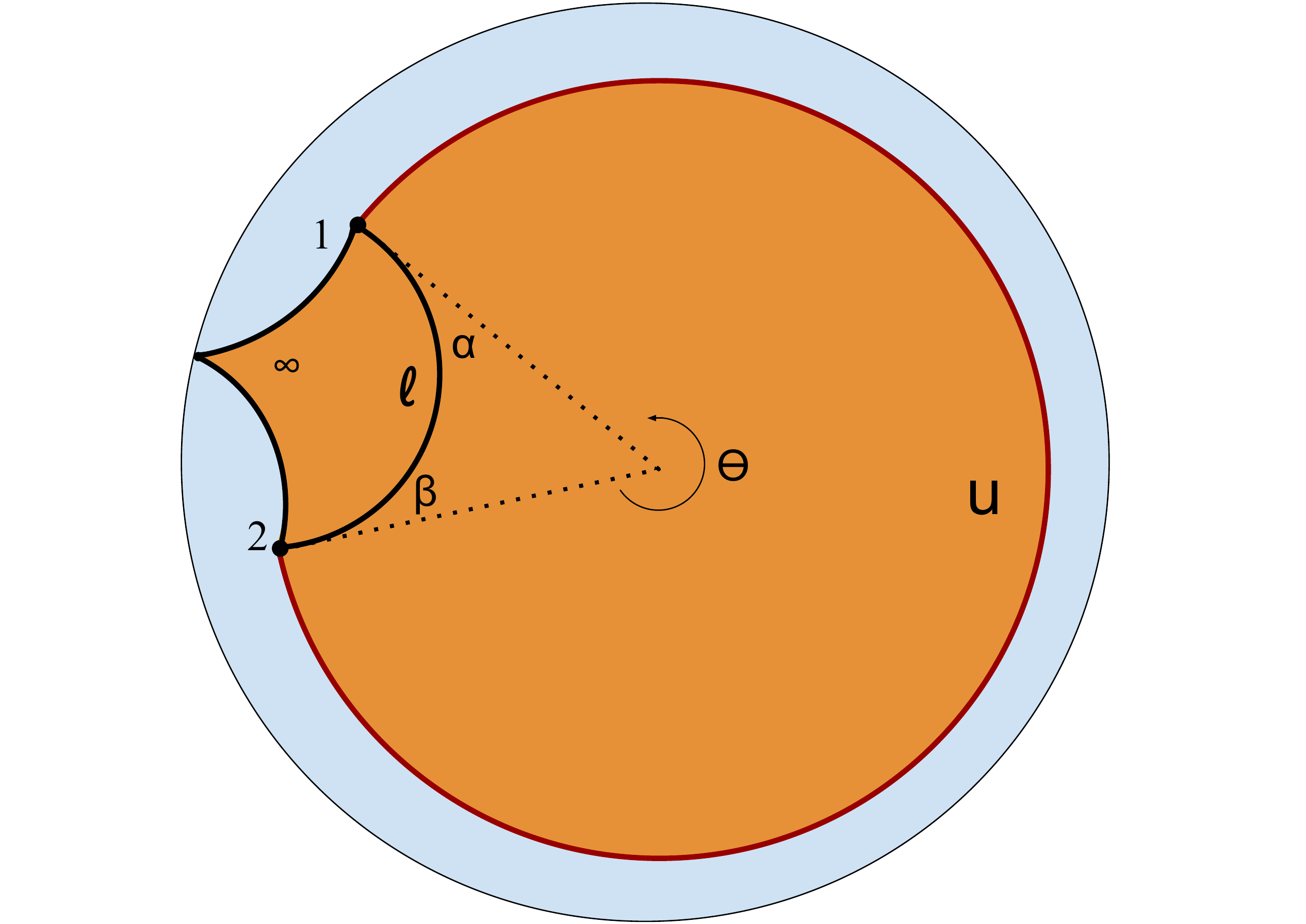}%
    \label{fig:propagator}%
  }%
  \caption{The singular ``phase" for different ordering and the geometric representation of the propagator} 
  \label{fig:Propagator}
\end{figure*}

To get a limit directly at the level of the mechanical system it is useful to define a  
rescaled 
coordinate, $z$, via 
\begin{equation}
    y=  z/q.  \la{yAndq}
\end{equation}
After taking the large $q$ limit, the boundary particle propagator becomes \footnote{see the Appendix \ref{appendix:large q} for details}: 
\bea\label{heat kernel}
G(u,\boldsymbol{x_1},\boldsymbol{x_2})&=&{ 1 \over q } e^{-2\pi q \theta(x_2-x_1) }\tilde{K}(u,\boldsymbol{x_1},\boldsymbol{x_2});~~~~~~~~~~~~~~~~~~~~~~~~~~~~~~~q\gg 1.\\
\tilde{K}(u,\boldsymbol{x_1},\boldsymbol{x_2})&=&e^{-2  \frac{z_1+z_2}{x_1-x_2} }\frac{2\sqrt{z_1z_2}}{\pi^2 |x_1-x_2|}\int_0^\infty ds s\sinh(2\pi s)e^{-\frac{s^2}{2 } u}K_{2is}(\frac{4 \sqrt{z_1z_2}}{|x_1-x_2|});~~~~~~~~
\la{HeatLim}\\
&=&e^{-2  \frac{z_1+z_2}{x_1-x_2} }\frac{\sqrt{2}}{\pi^{3/2} u^{3/2}}\frac{\sqrt{z_1z_2}}{ |x_1-x_2|}\int_{-\infty}^{\infty}d\xi (\pi+i\xi)e^{-2\frac{(\xi-i\pi)^2}{u}-\frac{4 \sqrt{z_1z_2}}{|x_1-x_2|}\cosh \xi}.\la{HeatLim2}
\eea
The original phase factor $e^{i \varphi(\boldsymbol{x_1},\boldsymbol{x_1})}$ factorizes into a product of singular ``phase" $e^{-2\pi q \theta(x_2-x_1)}$, with $\theta$ the step function, 
and a regular ``phase" $e^{-2 {z_1+z_2\over x_1-x_2}}$.  The singular ``phase" is the same order as the topological piece in (\ref{regularized action}). 
In order to have a finite result they should cancel between each other. This can only be satisfied if the $x_is$ are in cyclic order. 
As shown in figure (\ref{fig:phasefactor}), the product of singular ``phase" gives $-2\pi q$ for cyclic order $x_i$s and this would cancel with the topological action $2\pi q$.  
While for other ordering of the $x_i$s, this would have $-2 \pi n q$ for $n=2,3,...$ and is highly suppressed in the limit $q$ goes to infinity. 
This cyclic order is telling us where the interior of our space time is.  The magnetic field produces a preferred orientation for the propagator. 
 After fixing the order, all our formulas only depend on $\tilde{K}(u,\boldsymbol{x_1},\boldsymbol{x_2})$ which has no $q$ dependence. 
 The residual $q$ factor in \ref{heat kernel} cancels out the additional $q$ from the measure of coordinate integral, ${ dx dy \over y^2 } \to q { dx dz \over z^2 } $. 
In conclusion, after taking the limit we get a finite propagator equal to \nref{HeatLim}, which should be multiplied by a step function $ \theta( x_1 - x_2 ) $ that imposes the right order.

The final function $\tilde{K}(u,\boldsymbol{x_1},\boldsymbol{x_2})$ has the structure of $e^{-2{z_1+z_2\over x_1-x_2}}f(u,{z_1z_2\over (x_1-x_2)^2})$. This can be understood directly from the $SL(2)$ symmetry. 
After taking the large $q$ limit, the $SL(2,R)$ charges become
\begin{equation} \la{SLTwoLq}
    L_0=i (x\partial_x+z\partial_z);~~~~L_{-1}=i\partial_x;~~~~~~L_{1}=-ix^2\partial_x-2ix z\partial_z-2i z.
\end{equation}
We can check that they still satisfy the SL(2) algebra.
If we drop the last term in $L_1$,  the $SL(2,R)$ charges become the usual differential operators on $EAdS_2$.  And the propagator will have only dependence on the geodesic distance. When $L_1$ operator is deformed, the condition of $SL(2,R)$ invariance   fixes the structure of the propagator as follows. 
The $L_0$ and $L_{-1}$ charges are not deformed and they imply that the 
only combinations that can appear are 
\begin{equation}
  v \equiv  {z_1+z_2\over x_1-x_2} ~~~~~~~~~~~~~\text{and}~~~~~~~~~~~~~~ w\equiv {z_1 z_2\over(x_1-x_2)^2}.
\end{equation}
Writing the propagator as
$\tilde{K}(u,\boldsymbol{x_1},\boldsymbol{x_1})=k(v,w) $ 
and 
requiring it to be invariant under $L_1$ gives the following 
equation for $\alpha$:
\begin{equation}\label{phaseequation}
\partial_v k  +2  k =0  \longrightarrow k = e^{ - 2   v } h(w) 
\end{equation}
\begin{equation}
    \tilde{K}(u,\boldsymbol{x_1},\boldsymbol{x_2})=e^{-2 {z_1+z_2\over x_1-x_2}}f(u,{z_1 z_2\over(x_1-x_2)^2}),
\end{equation}
The full function can also be determined directly as follows. Again we impose the propagator equation (or heat equation) 
\bea
0 &=& \left[ \partial_u +  {1\over 2}\left( L_0^2+ { 1 \over 2} L_{-1}L_{1} +  { 1 \over 2} L_{1}L_{-1}  \right) -{1\over 8}  \right] \tilde K  
\cr
0 & = & [  { s^2 \over 2} + {w^2\over 2}\partial_w^2+2w +{1\over 8}] K_s(w) 
 \eea
where $L_a$ are given in \nref{SLTwoLq} and are acting only on the first argument of $\tilde K$. 
 The solution of the last equation which is regular at short distances ($w \to \infty$) is $\sqrt{w}$ times 
 the Bessel K 
 function in \nref{heat kernel}. 
 
 We can also directly determine the measure of integration for $s$ by demanding that the propagator at $u=0$ is
 a $\delta$ function or by demanding the propagator compose properly. This indeed is the case with  the $s\sinh{ 2 \pi s}$ function in \nref{HeatLim}. 
To explicitly show the above statement, it will be useful to use spectral decomposition of the propagator:
 \begin{equation}\label{SpectralDecomposition}
\tilde{K}(u,\boldsymbol{x_1},\boldsymbol{x_2})=\int ds {2s\sinh(2\pi s)\over \pi^3} e^{-{s^2u\over 2}}\int dk \sqrt{z_1 z_2} e^{ik(x_1-x_2)}K_{2is}(2\sqrt{2i k z_1})K_{2is}( 2\sqrt{2ikz_2}).
 \end{equation}
It can be easily checked that the special functions $f_{k,s}(x,z)=\sqrt{z}e^{ikx}K_{2is}( 2\sqrt{2ikz})$ are delta function normalizable eigenmodes of the large $q$ Hamiltonian:
\begin{equation}
\int {dx dz\over z^2}f_{k_1,s_1}f_{k_2,s_2}=\delta(k_1-k_2)\delta(s_1-s_2){\pi^3\over 2s\sinh (2\pi s)}
\end{equation}
Notice that the inner product fixes the integral measure completely in (\ref{SpectralDecomposition}), and the composition relation is manifestly true:
\begin{equation}
\int {dx dz\over z^2}\tilde{K}(u_1,\boldsymbol{x_1},\boldsymbol{x})\tilde{K}(u_2,\boldsymbol{x},\boldsymbol{x_2})=\tilde{K}(u_1+u_2,\boldsymbol{x_1},\boldsymbol{x_2})
\end{equation}
At short time the propagator has the classical behavior:
\begin{equation}
\tilde{K}(u,\boldsymbol{x_1},\boldsymbol{x_2})\sim \delta(x_1-x_2+u z_2)e^{-{(z_1-z_2)^2\over 2 u z_2}}
\end{equation}
This form of singularity is expected since we are taking the large $q$ limit first and thus the velocity in $x$ direction is fixed to be $z$. In the original picture of finite $q$ we are looking at the time scale which is large compare to AdS length but relatively small such that the quantum fluctuations are not gathered yet.

 The integral structure in the propagator (\ref{HeatLim}) has an obvious meaning: integrating over $s$ represents summing over all energy states with Boltzmann distribution $e^{-E u}$, and the Bessel function stands for fixed energy propagator. 
We want to stress that the argument in the Bessel function is unusual, and at short distance it approaches a funny limit:
\begin{equation}
K_{2is}(\frac{4 }{\ell})\simeq \sqrt{\pi \over 8\ell}e^{-\frac{4 }{\ell}},~~~~~\ell={|x_1-x_2|\over\sqrt{z_1 z_2}}\rightarrow 0.
\end{equation}
One should contrast this exponential suppression with the short distance divergence in QFT which is power law.
In our later discussion of exact correlation function with gravity backreaction, we will see that this effect kills UV divergence from matter fields. 

To obtain the expression (\ref{HeatLim2}), we use the integral representation for the Bessel function and the final result has some interesting physical properties: 

Firstly, we see that at large $u$ the time dependence and coordinate dependence factorized.  
So, at large time we have a universal power law decay pointed out in \cite{Bagrets:2016cdf}. 

Secondly, as we said before, the phase factor $e^{-2\frac{z_1+z_2}{x_1-x_2}}$ is equal to the (regularized) Wilson line $e^{-q\int_1^2 a}$ stretched along the geodesic connection between location $1$ and $2$ (Figure \ref{fig:propagator}).  The field $a$ depends on our choice of gauge, our convention corresponds to fix the minimum value of $a$ at infinity and then the Wilson line is equal to $e^{-q A}$, where $A$ is the area of a hyperbolic triangle spanned by $1,2$ and $\infty$. 

Thirdly, defining $2\pi+2i \xi$ as $\theta$, then $\theta$ has the meaning of the spanned angle at the horizon (Figure \ref{fig:propagator}). Then the gaussian weight $e^{-2{(\xi-i\pi)^2\over u}}=e^{\theta^2\over 2u}$ can be understood from the classical action along the boundary with fixed span angle $\theta$.  The boundary drawn in the figure represents a curve with fixed (regularized) proper length $u$ in $H_2$.

Lastly, the factor $e^{-\frac{4 \sqrt{z_1z_2}}{|x_1-x_2|}\cosh \xi}=e^{{4\cos\theta/2\over \ell}}$ is equal to $e^{-q(\alpha+\beta)}$, which is a corner term that arise from JT gravity in geometry with jump angles.  Here $\alpha$ and $\beta$ are defined as the angle spanned by the geodesic with fixed length and the ray coming from horizon to the boundary.

In summary the propagator can be understood as an integral of JT gravity partition functions over geometries \ref{fig:propagator} with different $\theta$s.

 Finally, let us comment on the issues we raised in section 4.2. In the large $q$ limit we are considering the propagator at relatively large distances and in a regime where locally in $AdS$ the integration over paths that fluctuate
 wildly is suppressed. Alternatively we can say that in the integration over paths we put a UV cutoff which 
 is large compared to $1/q$ but small compared to the AdS radius. This is the non-relativistic regime for the
 boundary particle. The quantum effects are still important at much longer distances due to the large size 
 of $AdS$. 
 In addition, if we have quantum fields in AdS, then their partition functions for these fluctuating contours that
 have fluctuations over distances larger than the $AdS$ radius are expected to depend on this shape in a local way. Due to the symmetries of $AdS_2$, this is simply expected to renormalize the action we already have without introducing extra terms. This can be checked explicitly for conformal field theories by using
 the conformal anomaly to compute the effective action of the CFT$_2$ on a portion of $H_2$ (Appendix \ref{appendix: CFT effective action}).

\subsection{Wheeler-DeWitt Wavefunction } 
\label{WdW}
In the pure JT theory we can think about quantizing the bulk theory and obtaining the 
Wheeler-DeWitt wavefunction. This was discussed in the classical limit by Harlow and Jafferis \cite{Harlow:2018tqv}. 

The Wheeler-DeWitt wavefunction can be created by Euclidean evolution of the boundary and hence is closely related to the propagator we have discussed above. The wavefunction in Lorentzian signature could then be obtained by analytic continuation of the boundary time. 
The Euclidean evolution can be specified by either of the two parameters: the proper length $u$ or energy $E$.  Choosing a different parameter corresponds to imposing a different boundary condition in JT theory. In general there are four possible choices of boundary conditions in 2d dilaton gravity, there are two sets of conjugate variables: $\lbrace \phi_b,K\rbrace$, and $\lbrace u,E\rbrace$ \footnote{Energy $E$ is proportional to the normal derivative of the dilaton field at the boundary.}.  In preparing the wavefunction we fix the boundary value of dilaton and hence there are only two choices of the parameter ($u$ or $E$).  We denote the corresponding wavefunction as $|u\rangle_G$ and $|E\rangle_G$ respectively. 
In terms of holographic considerations, $|u\rangle_G$ represents a thermofield double state: 
\begin{equation}
|u\rangle_G\sim\sum\limits_{n}e^{-E_n u}|E_n\rangle_L |\bar E_n\rangle_R
\end{equation}
and $|E\rangle_G $ is like an average of energy eigenstates in a window of energy $E$:
\begin{equation}\label{EnergyStateDefinition}
 |E\rangle_G\sim {1\over \delta E}\sum\limits_{ |E-E_n|<\delta E}|E_n\rangle_L |\bar E_n\rangle_R.
\end{equation}
 The width of the energy window is some coarse graining factor such that the summation contains $e^{S_0}$ states and does not show up clearly in gravity.\footnote{If one understand getting $|E\rangle$ state from integrating over thermofield double state in time direction, then a natural estimate on $\delta E$ is ${1\over T}$, where $T$ is the total time one integrate over.  The validity of JT description of boundary theory is $T<e^{S_0}$, and we get $\delta E>e^{-S_0}$. For $T>e^{S_0}$, there are other possible instanton contributions.  The proper gravitational theory at this regime is studied in paper \cite{Saad:2018bqo} }
	
\begin{figure}
\centering
\includegraphics[scale=0.5]{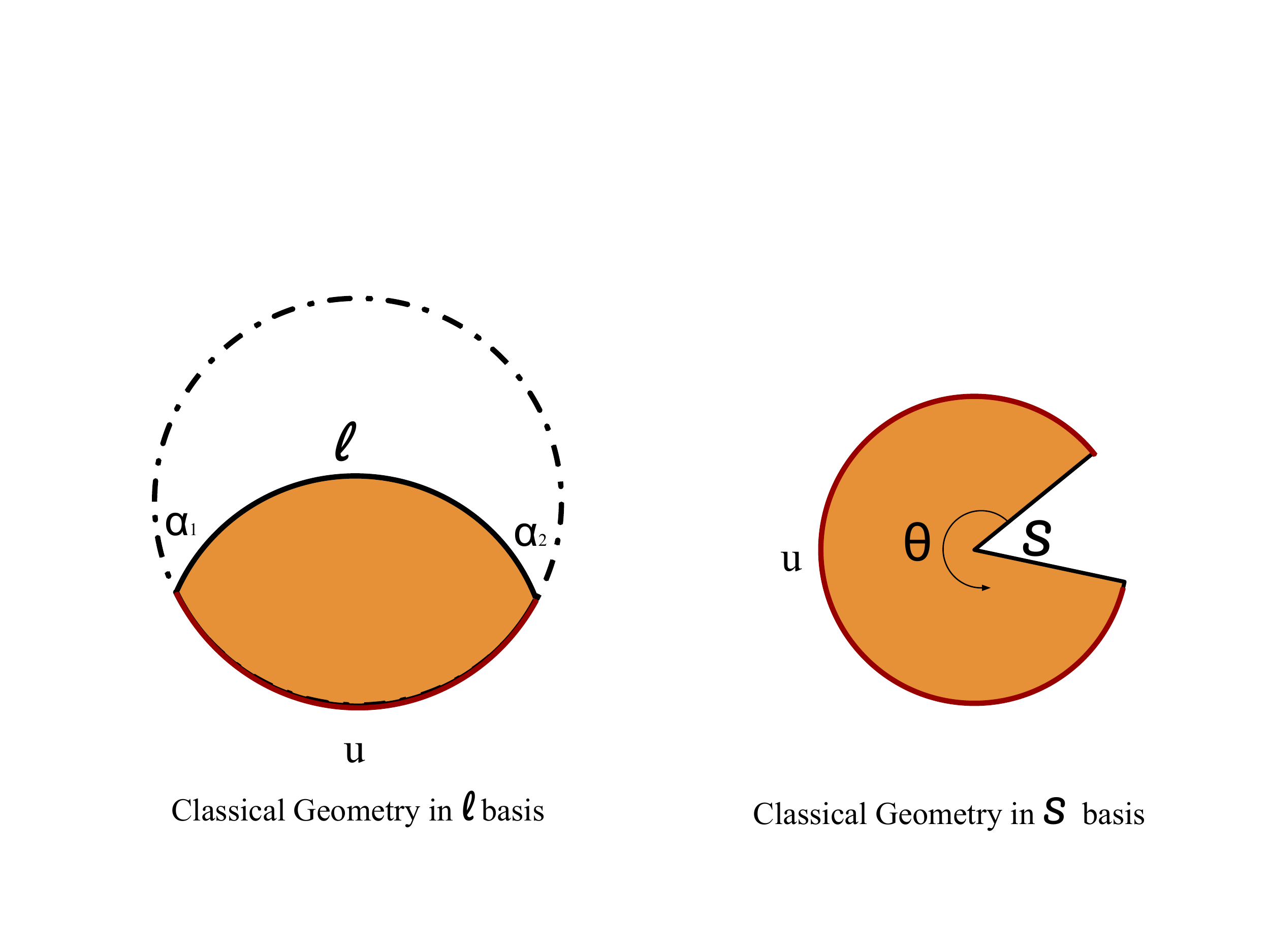}
\caption{Classical Geometry in $\ell$ and $S$ basis}
\label{fig:Pac-Man}
\end{figure}

With the definition of the states, one can evaluate them in terms of different basis.  There are three natural bases turn out to be useful, we call them $S$, $\eta$ and $\ell$ bases.  Basis $S$ corresponding to fix the horizon value of dilaton field $\phi_h$, or equivalently by Bekenstein-Hawking formula, the entropy of the system.  The canonical conjugate variable of $S$ will be called $\eta$ and that characterizes the boost angle at the horizon.  $\ell$ stands for fixing geodesic distance between two boundary points.  To see that the horizon value of the dilaton field is a gauge invariant quantity, one can do canonical analysis of JT gravity.   With ADM decomposition of the spacetime metric, one can get the canonical momenta and Hamiltonian constraints of the system \cite{LouisMartinez:1993eh}:
\begin{eqnarray}
ds^2&=&-N^2dt^2+\sigma^2(dx+N^xdt)^2;~~~~~~~~~~~~~~~~~~~~\\
    \mathcal{H}&=&-\Pi_{\phi}\Pi_{\sigma}+\sigma^{-1}\phi''-\sigma^{-2}\sigma'\phi'-\sigma\phi;~~~~\mathcal{H}_x=\Pi_{\phi}\phi'-\sigma\Pi_{\sigma}';\label{Hamiltonian Constraint}\\
    \Pi_{\phi}&=&N^{-1}(-\dot{\sigma}+(N^x\sigma)')=K\sigma;~~~~~~~~~~~~~\Pi_{\sigma}=N^{-1}(-\dot{\phi}+N^x\phi')=\partial_n\phi.\label{Canonical Momentum}
\end{eqnarray}
That is the dilaton field is canonically conjugate to the extrinsic curvature and boundary metric is canonical conjugate to the normal derivative of the dilaton field (both are pointing inwards). 
By a linear combination of the Hamiltonian constraints (\ref{Hamiltonian Constraint}), one can construct the following gauge invariant quantity $C$:
\begin{equation}
-{1\over\sigma}(\phi'\mathcal{H}+\Pi_{\sigma}\mathcal{H}_x)={1\over 2}(\Pi_{\sigma}^2+\phi^2-{\phi'^2\over \sigma^2})'\equiv C[\Pi_{\sigma},\phi,\sigma]'\sim 0
\end{equation}
The Dirac quantization scheme then tells us that the quantity $C$ has a constant mode which is gauge invariant (commute with Hamiltonian constraint).
Choosing the gauge that normal derivative of the dilaton is zero, we can solve the Hamiltonian constraint:
\begin{equation}
\phi^2-(\partial_{X}\phi)^2=2C\equiv S^2~~~~~\rightarrow~~~~~~\phi(X)=S\cosh X,
\end{equation}
where $dX= \sigma dx$ is the proper distance along the spatial slice.  Because the normal derivative of dilaton field is zero, the minimum value of dilaton at this spatial slice is actually a local extremum in both directions. Therefore, the minimal value of dilaton field, namely $S$, is a global variable. The classical geometry in this gauge is a ``Pac-Man" shape (right figure in figure \ref{fig:Pac-Man}).
Focusing on the intersection region of the spatial slice and the boundary, we have the spatial slice is orthogonal to the boundary.
This is because we are gauge fixing $\partial_n\phi=0$ on the spatial slice, and $\phi=\phi_b$ on the boundary. 
The ADM mass of the system, after regularization, is then $M=\phi_b(\phi_b-\partial_{X}\phi)$ \cite{Maldacena:2016upp}. Substituting the behavior of $\phi(X)$ we get:
\begin{equation}
M={S^2\over 2}.
\end{equation}
This is the same relation in \ref{energy relation} and therefore we can interpret the $s$ variable as entropy of our system $S$.

For the purpose of fixing geodesic distance, it is convenient to think of doing the path integral up to a slice $L$ with zero extrinsic curvature. This picks out a particular slice (left figure in Figure \ref{fig:Pac-Man}) among the solutions obeying the Hamiltonian constraint. 
The WdW wavefunction can be evaluated as an Euclidean path integral with fixed (rescaled) geodesic distance $d$ between the two boundary points:
\begin{equation}\label{WdWPath}
\begin{split}
	 \Psi(u;d)&=\int {\cal D}g{\cal D}\phi e^{{1\over 2}\int \phi(R+2)+\int_{L}\phi K+\phi_b\int_{Bdy}(K-1)+\phi_b(\alpha_1+\alpha_2-\pi)}=\int {\cal D }f e^{\phi_b\int_{Bdy} (K-1)+\phi_b(\alpha_1+\alpha_2-\pi)}\\
&=\int {\cal D}\boldsymbol x e^{-m \int_{Bdy}\sqrt{g}+q\int_{Bdy} a+q\int_{L} a+\pi q}
\end{split}
\end{equation}
Here we are fixing the total length of $L$ to be $d$ and the proper length of the boundary to be $u$. $\alpha_1$ and $\alpha_2$ in the last expression denotes the jump angle at the corner coming from the singular contribution of the extrinsic curvature and should be integrated over.  Without the $e^{q\int_L a}$ factor in (\ref{WdWPath}), the path integral corresponds to the propagator (\ref{HeatLim}).  Remember that the phase factor is equal to $e^{-q\int_L a}$, so the wavefunction in $\ell$ basis is actually the propagator (\ref{HeatLim}), with the phase factor stripped out 
\be\label{wavefunction}
\Psi(u; \ell)\equiv \langle \ell|u\rangle_G =  \frac{2}{\pi^2 \ell}\int_0^\infty ds s\sinh(2\pi s)e^{-\frac{s^2}{2 } u}K_{2is}(\frac{4 }{\ell}) ,~~~~~~~~\ell={|x_1-x_2|\over\sqrt{z_1 z_2}}.
\ee
$\ell$ is a function of the regularized geodesic distance $d$ between $\boldsymbol{x_1}$ and $\boldsymbol{x_2}$ : $\ell=e^{d\over 2}$.
The semiclassical of $\Psi(u;\ell)$ can be obtained using formula (\ref{HeatLim2}), in the exponent we get saddle point result:
\begin{equation}
\Psi(u;\ell)\sim \exp[-{2(\xi_*-i\pi)^2\over u}+{4\over u}{\xi_*-i\pi\over \tanh \xi_*}];~~~~~~~~{\xi_*-i\pi\over u}=-{\sinh \xi_*\over \ell}.
\end{equation}
The same saddle point equation and classical action was obtained in \cite{Harlow:2018tqv} by a direct evaluation in JT gravity.

The wavefunction with fixed energy boundary condition can obtained by multiplying $\Psi(u;\ell)$ by 
 $e^{ + E u} $ and integrating over $u$ along the imaginary axis. This sets $E = {s^2 \over 2} $
 in the above integral over $s$. So this wavefunction has a very simple expression:
 \be\label{fixed energy wavefunction}
 \Psi(E;\ell)\equiv \langle \ell|E\rangle_G = \rho(E)\frac{4}{ \ell} K_{i \sqrt{ 8 E} }(\frac{4 }{\ell}).~~~~~~~~
\ee
The classical geometry for $\Psi(E;\ell)$ is the same as the left figure in Figure \ref{fig:Pac-Man}, with fixing energy on the boundary.
We want to stress that it is important to have the $\rho(E)$ factor in (\ref{fixed energy wavefunction}) for a classical geometry description since we are averaging over the states.
We can roughly think of ${4\over \ell}K_{i\sqrt{8E}}({4\over \ell})$ as a gravitational ``microstate" $|\mathcal{E}\rangle$ with fixed energy $E$.  Such a ``microstate" will not have a classical geometry representation and therefore is just a formal definition. 
The inner product between wavefunctions is defined as $\langle\Psi_1|\Psi_2\rangle=\int_0^{\infty}d\ell \ell \Psi_1^*(\ell)\Psi_2(\ell)$.

\begin{figure}
\centering
\includegraphics[scale=0.4]{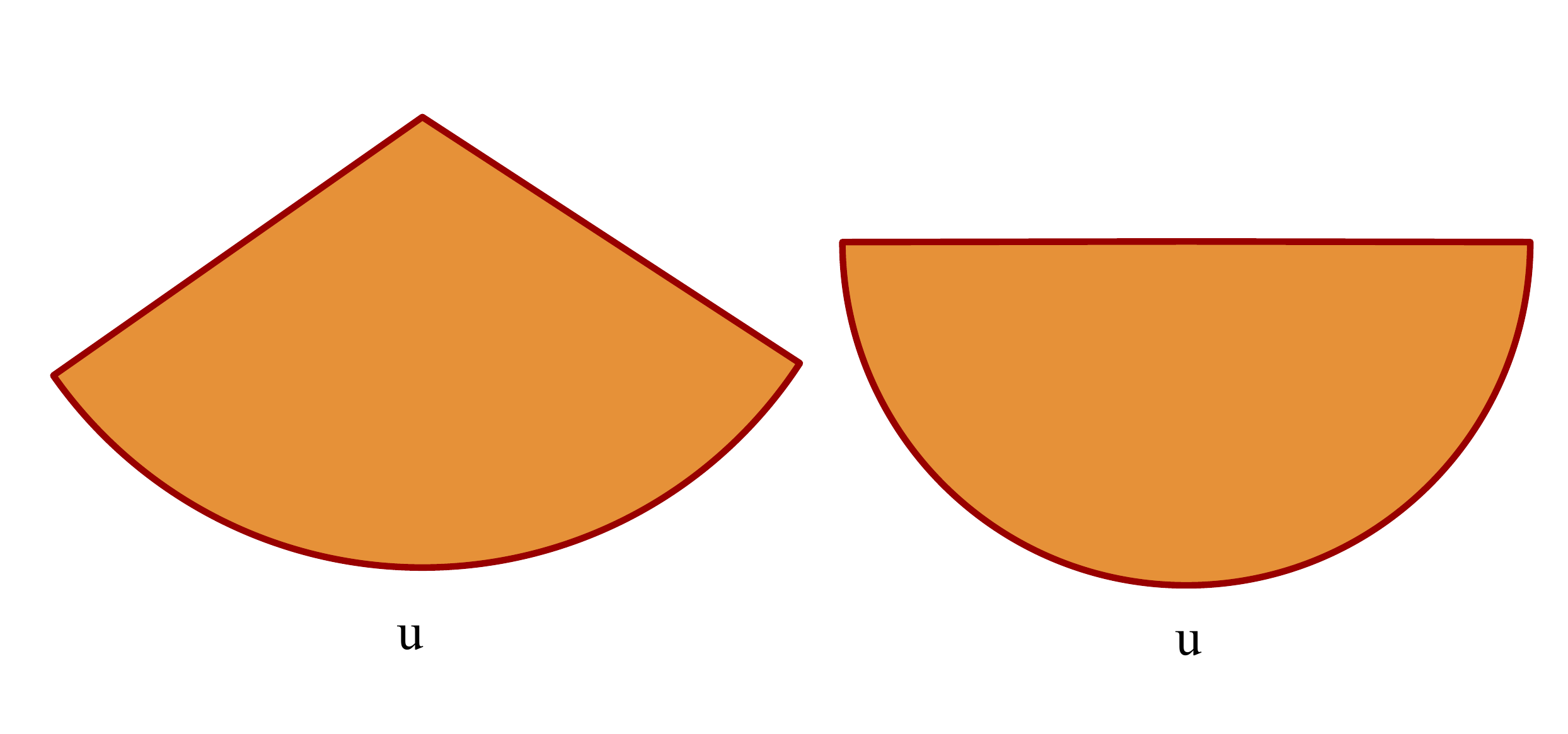}
\caption{Euclidean geometries with different cusps.}
\label{fig:Boostangle}
\end{figure}

Going to the entropy basis $S$, it is easy to start with $\Psi(E)$.
Because of the identity $E={S^2\over 2}$, expanding $\Psi(E)$ in the $S$ basis is diagonal:
\begin{equation}
\Psi(E;S)\equiv \langle S|E\rangle_G=\sqrt{\rho(S)}\delta(E-{S^2\over 2})
\end{equation}
We put this square root of $\rho(S)$ factor in the definition of $S$ basis such that inner product between different $S$ state is a delta function $\langle S|S'\rangle=\delta(S-S')$.  This factor is also required such that the classical limit matches with gravity calculation. 
Integrating over energy with Boltzman distribution, we can get the expression of thermofield double state in the $S$ basis:
\begin{equation}
\Psi(u;S)\equiv \langle S|u\rangle_G=\int dE  e^{-u E}\langle S|E\rangle_G=\sqrt{\rho(S)}e^{-{uS^2\over 2}}
\end{equation}
In the semiclassical limit, the wavefunction becomes gaussian and coincides with the on shell evaluation of the ``Pac-Man" geometry (Figure \ref{fig:Pac-Man}):
\begin{equation}\label{PsiS}
\Psi(u,S)\sim \sqrt{S}e^{\pi S -{u S^2\over 2}}
\end{equation}
The on shell calculation is straightforward: JT action in this geometry contains two parts: the Schwarzian action $\int(K-1)$ on the boundary and a corner contribution at the center: $S(\pi-\theta)$, where $\theta$ is the span angle at the horizon (Figure \ref{fig:Pac-Man}).  The Schwarzian action simply gives $E u={S^2 u\over 2}$ by direct evaluation. We can determine $\theta$ from $u$ since they are related with redshift: $\theta=u S$. Therefore the corner term gives: $\pi S-S^2 u$. Adding them up then gives us the classical action. 
We can also expand $S$ in terms of the $\ell$ basis, and relate $\Psi(\ell)$ with $\Psi(S)$ by a change of basis:
\begin{equation}
\langle\ell|S\rangle=\sqrt{\rho(S)}{4\over \ell}K_{2iS}({4\over \ell});~~~~~\Psi(E;\ell)=\int_0^{\infty} dS \langle \ell|S\rangle \langle S|E\rangle_G;
\end{equation}
Before discussing our last basis, we want to stress the simplicity of the wavefunction in $S$ basis (\ref{PsiS}) and the Gaussian factor resembles an ordinary particle wavefunction in momentum basis. We introduce our last basis $\eta$ as canonical conjugate variable of $S$, with an analog of going to position space of the particle picture in mind:
\begin{eqnarray}
|\eta\rangle &=&\int_0^{\infty} dS \cos(\eta S)|S\rangle;~~~~~~~~~~~~~\langle \ell|\eta\rangle=\int_0^{\infty} dS\cos(\eta S)\sqrt{\rho(S)}K_{2iS}({4\over \ell}){4\over \ell};~~~~~\\
\Psi(E,\eta) &=&\scaleto{\sqrt{\sinh(2\pi \sqrt{2E})\over 2\pi^2\sqrt{2E}}\cos(\eta \sqrt{2E})}{30pt};~~~~~\Psi(u,\eta)=\int_0^{\infty} dS \sqrt{\rho(S)}\cos(\eta S)e^{-{uS^2\over 2}}.~~~~
\end{eqnarray}
To understand the meaning of $\eta$ better, we can look at the classical behavior of $\Psi(u;\eta)$:
\begin{equation}
\Psi(u;\eta)\sim {1\over u}\exp[{\pi^2\over 2 u}-{\eta^2\over 2 u}]\left(e^{i\eta\pi\over u}\sqrt{\pi+i\eta}+e^{-{i\eta\pi\over u}}\sqrt{\pi-i\eta}\right)
\end{equation}
When $u$ is real, the wavefunction is concentrated at $\eta=0$ and has classical action of a half disk in the exponent.
When $u={\beta\over 2}+it$ which corresponds to the case of analytically continuing into Lorentzian signature, the density of the wavefunction $|\Psi(u,\eta)|^2$ is dominated by:
\begin{equation}
 |\Psi({\beta\over 2}+it,\eta)|^2\sim {\sqrt{\pi^2+\eta^2}\over \beta^2+4t^2}\exp[{2\pi^2\over \beta}]\left(\exp[-{2\beta(\eta-{2\pi t\over \beta})^2\over \beta^2+4t^2}]+\exp[-{2\beta(\eta+{2\pi t\over \beta})^2\over \beta^2+4t^2}]\right)
\end{equation}
showing the fact that $\eta$ is peaked at the Rindler time ${2\pi t\over \beta}$. We can therefore think of fixing $\eta$ as fixing the IR time or the boost angle at the horizon. 
The classical intuition for the boost angle is most clear in Euclidean geometry, where for fixed boundary proper time there can be different cusps at the horizon (Figure \ref{fig:Boostangle}).

\begin{figure}
\centering
\includegraphics[scale=0.4]{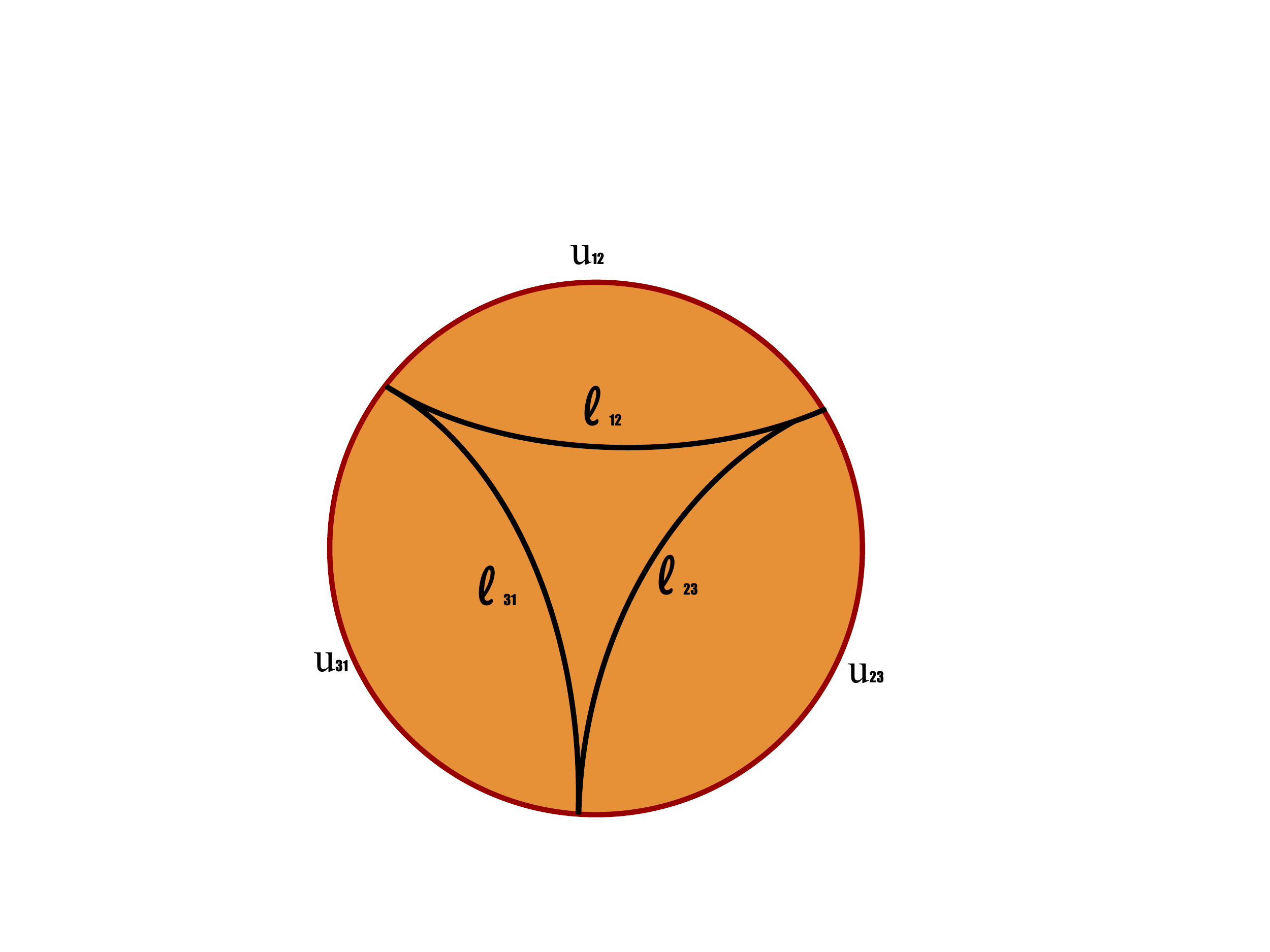}
\caption{Partition function from inner product of three wavefunctions.}
\label{fig:threewavefunction}
\end{figure}

One application of those wavefunctions is that we can take an inner product and get the partition function. 
However, there are also other ways to get the partition function. 
For example, we can concatenate three propagators and integrate over their locations.  This also gives the partition function by the composition rule of propagator. 
By the relation between propagator and wavefunction, we can also view this as taking an inner product of three wavefunctions with an interior state as in figure \ref{fig:threewavefunction}, where the interior state can be understood as an entangled state for three universes.
To be more precise, we can view the wavefunction as the result of integrating the bulk up to the geodesics with zero extrinsic curvature. Then the interior state is given 
by the area of the hyperbolic triangle in figure \ref{fig:threewavefunction}. 
The path integral for the hyperbolic triangle (denoted as $I(\ell_{12},\ell_{23},\ell_{31})$, where $\ell_{ij}={|x_i-x_j|\over \sqrt{z_iz_j}}$),  is a product of three phase factors,  which satisfies a nontrivial equality (with ordering $x_1>x_2>x_3$):
\begin{equation}\label{GHZ}
   \scaleto{ I(\ell_{12},\ell_{23},\ell_{31})=e^{-2({z_1+z_2\over x_1-x_2}+{z_2+z_3\over x_2-x_3}+{z_3+z_1\over x_3-x_1})}={16\over\pi^2}\int_0^{\infty}d\tau \tau \sinh (2\pi \tau) K_{2i\tau}({4\over \ell_{12}})K_{2i\tau}({4\over \ell_{23}})K_{2i\tau}({4\over \ell_{31}}).}{26pt}
\end{equation}
Recalling that the Bessel function represents the fixed energy ``microstate" $|\mathcal{E}\rangle$ (\ref{fixed energy wavefunction}) and ${\tau\over 2\pi^2} \sinh(2\pi \tau)$ is the density of state, this formula tells us that the interior state is a GHZ state for three universe:
\begin{equation}
I_{123}\sim\sum_n|\mathcal{E}_n\rangle_1|\mathcal{E}_n\rangle_2|\mathcal{E}_n\rangle_3.
\end{equation}
$I$ can also been viewed as a scattering amplitude from two universes into one universe. It constrains the SL(2,R) representation of the three wavefunctions to be the same.\footnote{Some thing similar happens for 2d Yang-Mills theory \cite{Witten:1992xu,Cordes:1994fc,Luca}.}
We can write down the partition function as:
\begin{equation}
Z_{JT}=\int_0^{\infty} \prod\limits_{\lbrace i j \rbrace \in \lbrace 12,23,31\rbrace}d\ell_{ij} \Psi(u_{12},\ell_{12})\Psi(u_{23},\ell_{23})\Psi(u_{31},\ell_{31})I_{\ell_{12},\ell_{23},\ell_{31}}.
\end{equation}
This same result also holds if we repeat the process $n$ times. It is interesting that we can view the full disk amplitude in these various ways.  

One can also extend our analysis to include matter field. One type of such wavefunction can be created by inserting operator during Euclidean evolution, and is analysed in appendix B. Note that because of the SL(2,R) symmetry is a gauge symmetry, our final state has to be a gauge singlet including matter field.

 \section{Correlation Functions in Quantum Gravity}
 
\subsection{Gravitational Feynman Diagram}

The propagator enables us to ``dress'' quantum field theory correlators to produce quantum gravity ones. 
Namely, we imagine that we have some quantum field theory in $H_2$ and we compute correlation functions of
operators as we take the points close to the boundary where they take the form 
\begin{equation}
    \langle O_1(\boldsymbol{x_1})...O_n(\boldsymbol{x_n})\rangle_{\rm QFT} = q^{ - \sum \Delta_i }z_1^{\Delta_1}..z_n^{\Delta_n}\langle O_1(x_1)...O_n(x_n)\rangle_{CFT}
\end{equation} 
The factor of $q$ arises   from \nref{yAndq}, and the last factor is simply defined as the    function 
that results after extracting the $z$ dependence. For example, for a two point function we get 
\begin{equation}
    \langle O_1(\boldsymbol{x_1})O_2(\boldsymbol{x_2})\rangle_{\rm QFT}=  
    q^{ - 2\Delta } z_1^{\Delta} z_2^{\Delta}{1\over |x_1-x_2|^{2\Delta}}.
\end{equation}
 
We can now use the propagator \nref{HeatLim} to couple the motion of the boundary and thus obtain the full 
quantum gravity expression for the correlator. The factors of $q$ are absorbed as part of the renormalization 
procedure for defining the full quantum gravity correlators. 
In this way we obtain 
\begin{figure*}[h!]
  \centering
  \subfigure[Witten Diagram]{%
    \includegraphics[width=0.3\textwidth]{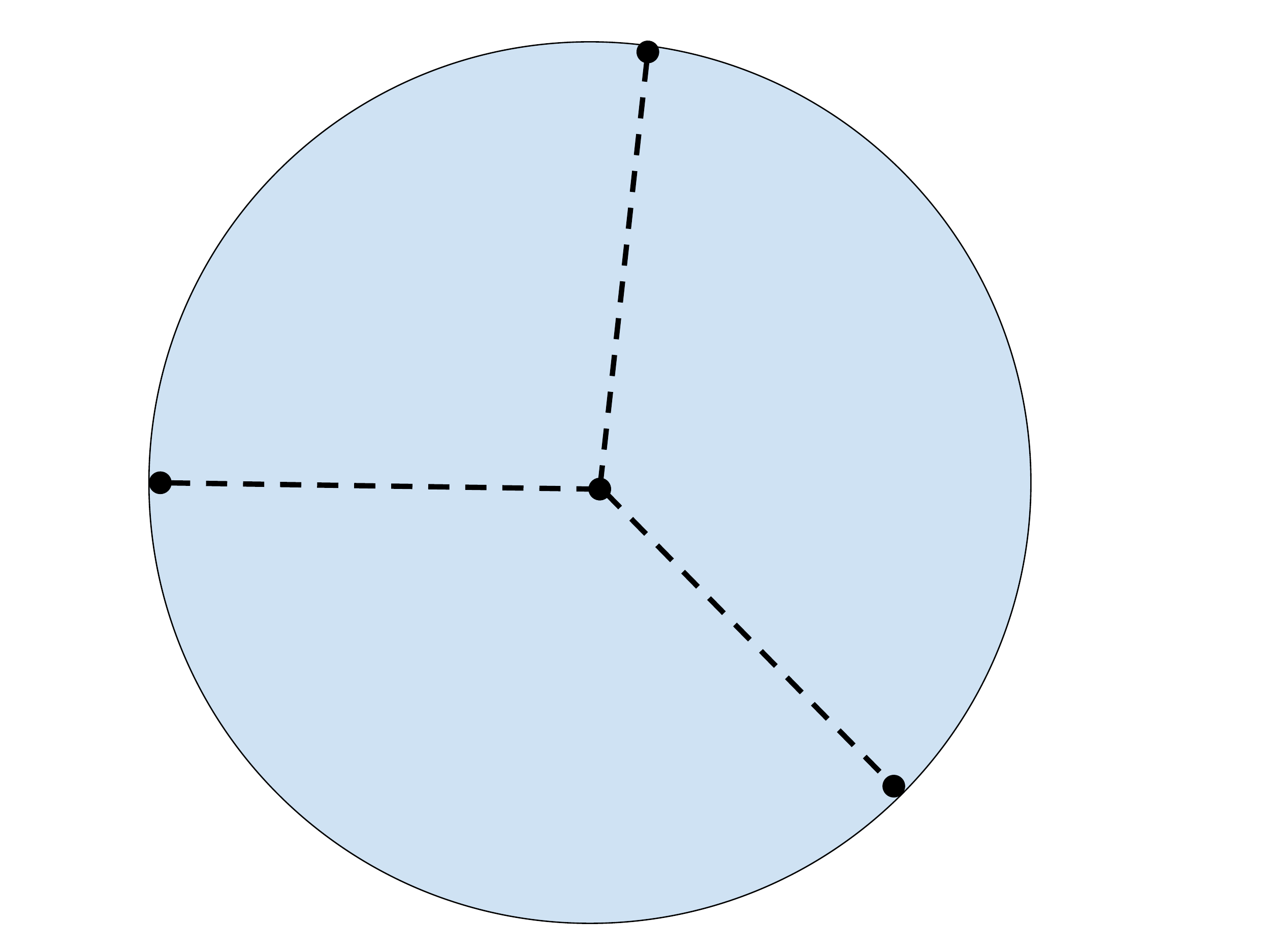}%
    \label{fig:Witten Diagram}%
    }\hspace{4cm}
    \subfigure[Gravitational Feynman Diagram]{%
    \includegraphics[width=0.3\textwidth]{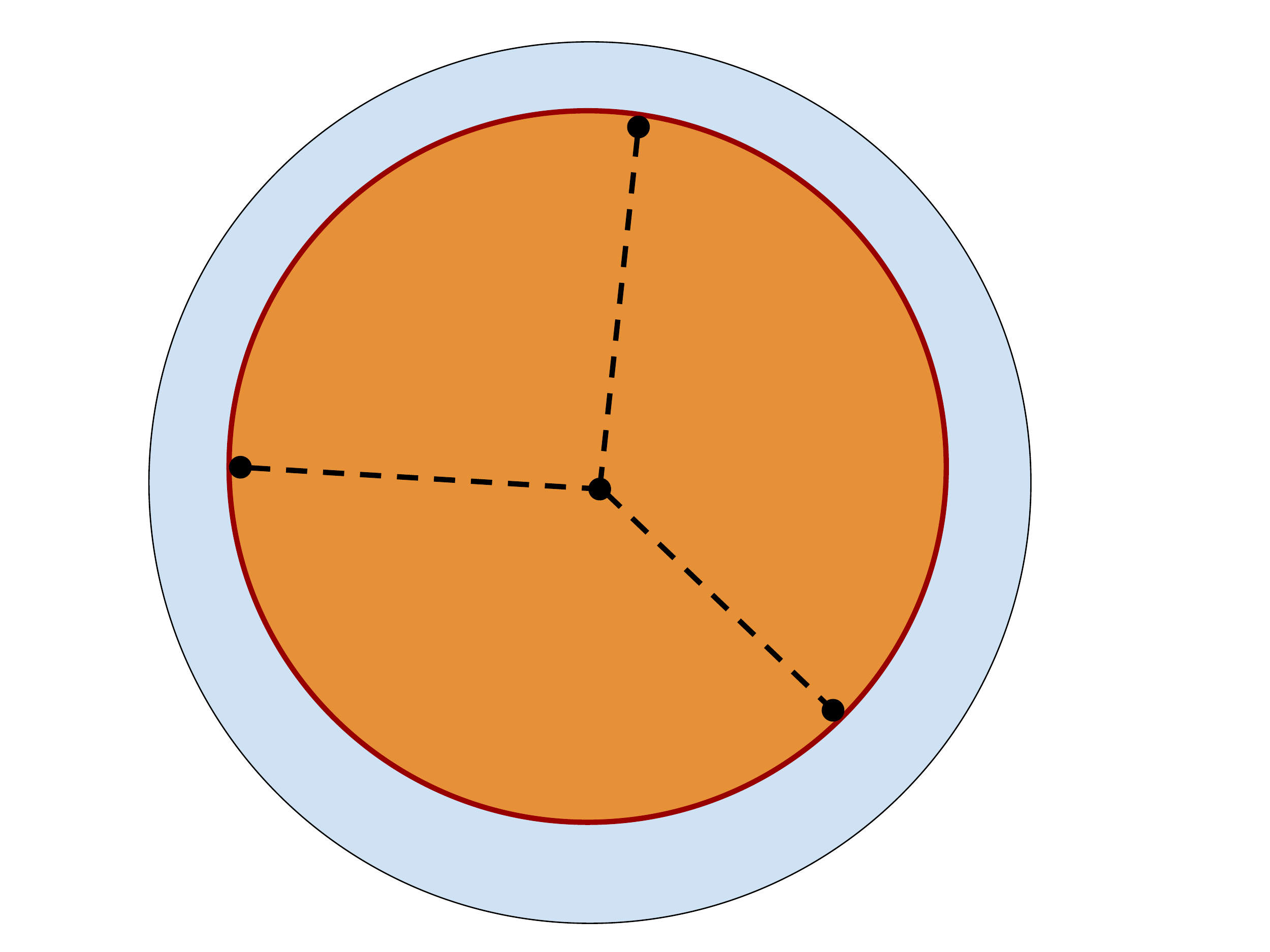}%
    \label{fig:Dynamical Witten Diagram}%
  }%
  \caption{Summation of ${1\over N}$ effects fluctuates the boundary of Witten Diagram} 
  \label{fig:witten diagram}
\end{figure*}
\bea\label{Higher Point Function}
\small{\langle O_1(u_1)...O_n(u_n)\rangle_{\text{\rm QG}}}&=&e^{2\pi q}\int {\prod_{i=1}^n{dx_i dy_i\over y_i^2}\over \text{V(SL(2,R))}} G(u_{12},\boldsymbol{x_1},\boldsymbol{x_2})G(u_{23},\boldsymbol{x_2},\boldsymbol{x_3})...G(u_{n1},\boldsymbol{x_n},\boldsymbol{x_1})\times 
\cr
& ~& \times \langle O_1(\boldsymbol{x_1})...O_n(\boldsymbol{x_n})\rangle_{\rm QFT} q^{ \sum_ i \Delta_i }
\eea
where the left hand side is the full quantum gravity correlator by definition. The last factor is the usual 
renormalization necessary to get something finite. 

The factor of $e^{ 2 \pi q } $ cancels with the  $q$ dependent ``phase'' factors in \nref{heat kernel} to give one if we order
the points cyclically ($x_1>x_2...>x_n$). This requires that we define more carefully the last propagator 
$G(u_{n1},\boldsymbol{x_n},\boldsymbol{x_1})$ as:
\be
e^{-2\pi q}\tilde K(u_{n1},\boldsymbol{x_n},\boldsymbol{x_1})=e^{-2\pi q}e^{-2 \frac{z_n+z_1}{x_n-x_1}}\frac{2\sqrt{z_nz_1}}{\pi^2 |x_n-x_1|}\int_0^\infty ds s\sinh(2\pi s)e^{-\frac{s^2}{2} u_{n1}}K_{2is}(\frac{4\sqrt{z_nz_1}}{|x_n-x_1|}).
\ee

The factor ${1\over \text{V(SL(2,R))}}$ in \nref{Higher Point Function} means that we should fix the $SL(2,R)$ gauge symmetry (Appendix \ref{gauge fix}). 
 
 In the end we can write down an expression where we have already taken the $q \to \infty $ limit
 \begin{equation}\label{final formula}
  \boxed{\scaleto{\langle O_1(u_1)...O_n(u_n)\rangle_{\text{QG}}=\int_{ x_1>x_2..>x_n}{\prod_{i=1}^n{dx_i dz_i}\over \text{V(SL(2,R))}} \tilde{K}(u_{12},\boldsymbol{x_1},\boldsymbol{x_2})...\tilde{K}(u_{n1},\boldsymbol{x_n},\boldsymbol{x_1})z_1^{\Delta_1-2}..z_n^{\Delta_n-2}\langle O_1(x_1)...O_n(x_n)\rangle_{\rm CFT}.}{26pt}}
 \end{equation}

This is one of the main results of our paper and it gives a detailed expression for correlation function in 2 dimensional quantum gravity in terms of the correlation functions of the QFT in hyperbolic space, or $AdS_2$. 

Notice that in usual $AdS/CFT$ the correlators  $\langle O_1(x_1)...O_n(x_n)\rangle_{\rm CFT}$ are an approximation to the full answer. This is sometimes computed by Witten diagrams.  
We get a better approximation by integrating over the metric fluctuations. 
In this case,  the non-trivial gravitational mode is captured by the boundary propagator. The formula \nref{final formula} includes all the effects of quantum gravity in the JT theory (in the Schwarzian limit). 
The final diagrams consist of the Witten diagrams for the field theory in $AdS$ plus the propagators for the boundary particle and we can call them ``Gravitational Feynman Diagrams", see figure \ref{fig:witten diagram}.

\subsection{Two Point Function}

Using formula \nref{final formula}, we can study gravitational effects on bulk fields such as its two point function:
\bea
\langle O_1(u)O_2(0)\rangle_{QG}=
   \includegraphics[width=50mm,trim=0 9cm 0 0]{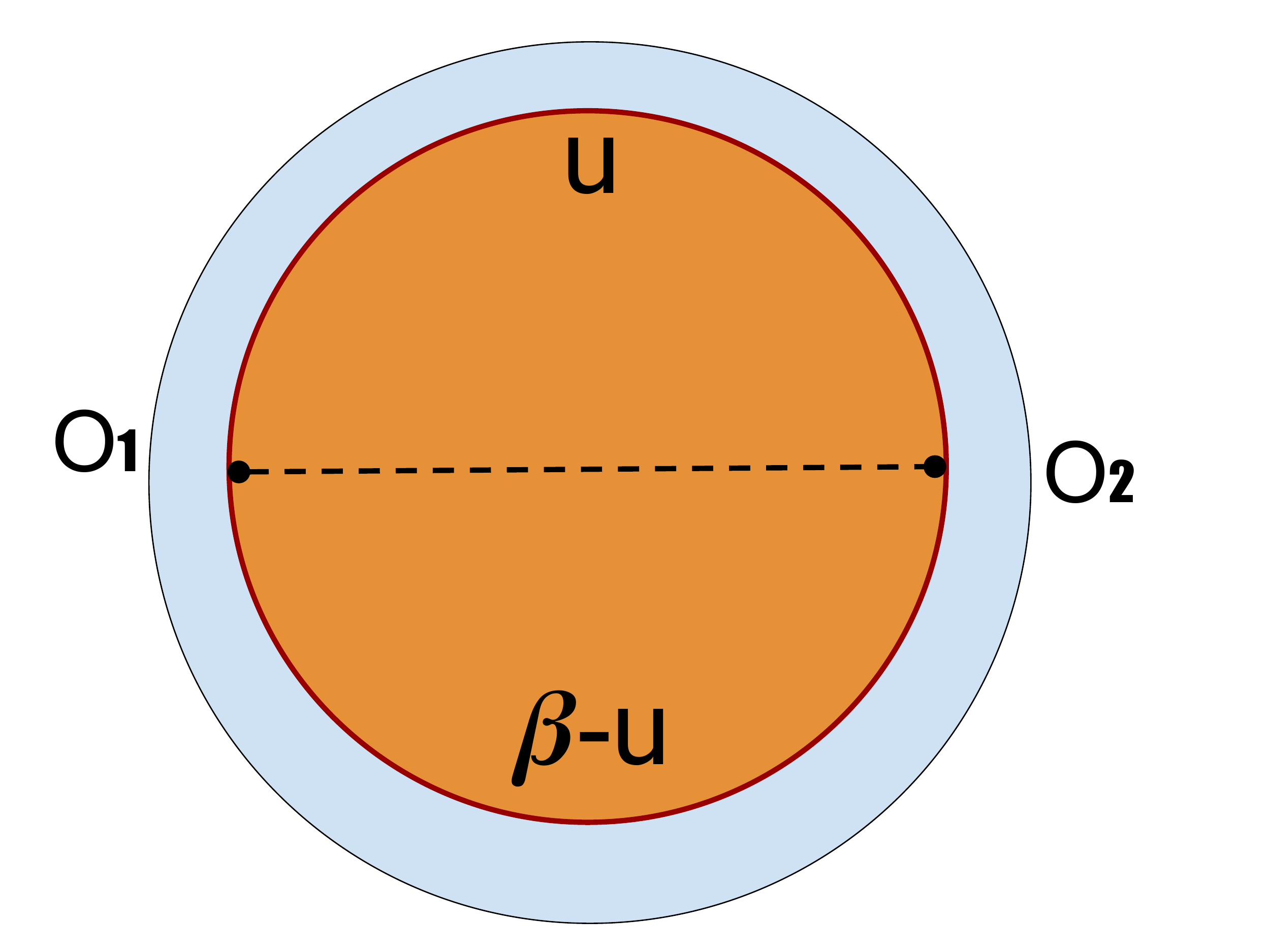}\\
   ~~~~~~~~~~~\nonumber\\
   ~~~~~~~~~\nonumber
\eea
The explicit expression for $\langle O_1(u)O_2(0)\rangle_{QG}$ with dimension $\Delta$ at temperature ${1\over \beta}$ is 
\footnote{We will just keep the $\Delta$ dependent constant 
since at last we will normalized with respect of partition function which corresponds to set $\Delta=0$.}:
\begin{equation}
  \scaleto{{1\over \text{V(SL(2,R))}}\int_{x_1>x_2} {dx_1 dx_2 dz_1 dz_2 \over z_1^2 z_2^2}\int_0^{\infty}ds_1 ds_2 \rho(s_1)\rho(s_2)e^{-{s_1^2\over 2}u-{s_2^2\over 2}(\beta-u)}K_{2is_1}({4\sqrt{z_1z_2}\over |x_1-x_2|})K_{2is_2}({4\sqrt{z_1z_2}\over |x_1-x_2|})({\sqrt{z_1z_2}\over |x_1-x_2|})^{2\Delta+2}}{25pt}.
\end{equation}
 To fix the $SL(2,R)$ gauge, we can choose $z_1=z_2=1$ and $x_2=0$. 
 Then the integral over $H_2$ space is reduced to a single integral over $x_1$,
 with a Jacobian factor $2x_1$ (Appendix \ref{gauge fix}):  
\be\label{space integral}
\int_0^{\infty}ds_1ds_2\rho(s_1)\rho(s_2)e^{-{s_1^2\over 2}u-{s_2^2\over 2}(\beta-u)}\int_0^{\infty} dx_1 ({1\over x_1})^{2\Delta+1}K_{2is_1}({4\over x_1})K_{2is_2}({4\over x_1}). 
\ee
the last integral can be interpreted as a matrix element of two point operator $O_1O_2$ between states $|E_1,\psi\rangle$ and $|E_2,\psi\rangle$, 
where $|\psi\rangle$ is the wavefunction of quantum field theory
and $|E\rangle_G$ represents the fixed energy gravitational state.
Integrating over $x$ can be thought as integrating over a particular gravitational basis, and we can see that the gravity wavefunction
suppress the UV contributions from quantum field theory ($K_{2is}({4\over x})\sim \sqrt{\pi x\over 8}e^{-4/x}$ for $x\sim 0$).
The final expression for the two point function is:
\bea\label{2point function}
\langle O_1(u) O_2(0)\rangle_{QG}={1\over \mathcal{N}}\int ds_1ds_2\rho(s_1)\rho(s_2)e^{-{s_1^2\over 2}u-{s_2^2\over 2}(\beta-u)}\frac{|\Gamma(\Delta-i(s_1+s_2))\Gamma(\Delta+i(s_1-s_2))|^2}{2^{2\Delta+1} \Gamma(2\Delta)};~~\\
={1\over \mathcal{N}}{\Gamma(2\Delta)\over u^{3/2} (\beta-u)^{3/2} 2^{4\Delta+4}\pi^3} \int_{c-i\infty}^{c+i\infty} d\theta_1d\theta_2 \theta_1\theta_2 e^{{\theta_1^2\over 2u}+{\theta_2^2\over 2 (\beta-u)}}{1\over (\cos{\theta_1\over 2}+\cos{\theta_2\over 2})^{2\Delta}}.~~~~~
\eea
In the second expression we write the integral in terms of variable $\theta$ using the second integral representation of the propagator (\ref{HeatLim2}).
The normalization constant can be determined by taking the $\Delta=0$ limit: $\mathcal{N}=Z_{JT}$. 

If we contemplate the result (\ref{2point function}) a little bit, then we find that the two integrals of $s_1$ and $s_2$ just represent the spectral decomposition of the two point function.  Indeed, under spectral decomposition we have $\langle O(u)O(0)\rangle=\sum\limits_{n,m}  e^{-E_n u-E_m (\beta-u)}|\langle E_n|O|E_m\rangle|^2$.  Compare with (\ref{2point function}), we can read out the square of matrix element of operator $O$:
\begin{equation}
    \scaleto{_G\langle E_1|O_L O_R|E_2\rangle_G=\delta E^{-2}\sum\limits_{\substack{|E_n-E_1|<\delta E\\|E_m-E_2|<\delta E}}|\langle E_n|O|E_m\rangle|^2=\rho(E_1)\rho(E_2){|\Gamma(\Delta-i(\sqrt{2E_1}+\sqrt{2E_2}))\Gamma(\Delta+i(\sqrt{2E_1}-\sqrt{2E_2}))|^2\over 2^{2\Delta+1}\Gamma(2\Delta)}.}{38pt}
\end{equation}
Remember the notation is that $|E\rangle_G$ stands for a gravitational state with energy $E$ and $|E_n\rangle$ stands for one side microstate (\ref{EnergyStateDefinition}).
We have put the measure $\rho(E)={1\over 2\pi^2}\sinh(2\pi\sqrt{2 E})$ in the definition of matrix element for the reason that in gravity it is more natural to consider an average of energy states as a bulk state.
To understand this formula a little bit better, we can consider the classical limit, namely large $E$.  In this limit the matrix element squared can be approximated as a nonanalytic function:
\begin{equation}
 _G\langle E_1|O_L O_R|E_2\rangle_G\propto |E_1-E_2|^{2\Delta-1}e^{2\pi \text{min}(\sqrt{2E_1},\sqrt{2E_2})}.
\end{equation}
  If we fix $E_1$ and varying $E_2$ from $0$ to infinity, the matrix element changes from $|E_1-E_2|^{2\Delta-1}\rho(E_2)$ to $|E_1-E_2|^{2\Delta-1}\rho(E_1)$ after $E_2$ cross $E_1$.  We can understand this behavior qualitatively as a statistical effect: the mapping from energy subspace $E_1$ to $E_2$ by operator $O$ is surjective when the Hilbert space dimension of $E_2$ is less that $E_1$ and is injective otherwise.
Another understanding is the following: the two point function is finite in a fixed energy state $|E_n\rangle$, which means the following summation of intermediate states $|E_m\rangle$ is order one: $\sum\limits_{m}|\langle E_n|O|E_m\rangle|^2$.  Looking at the case $E_m>E_n$, because of the density of states grows rapidly, the matrix element squared has to be proportional to ${1\over \rho(E_m)}$ to get a finite result. Multiplied by $\rho(E_n)\rho(E_m)$, we have $\rho(E_n)\rho(E_m)|\langle E_n|O|E_m\rangle|^2\sim  \rho(\text{min}(E_n,E_m))$.

 \subsection{ETH and the KMS condition} 
 
The Eigenstate Thermalization Hypothesis (ETH) is a general expectation for chaotic system. It expresses that the operator expectation value in an energy eigenstate can be approximated by thermal expectation value with effective temperature determined from the energy.  Such hypothesis can be tested with the knowledge of operator matrix elements. The two point function in microcanonical essemble is:
\begin{eqnarray}
    {1\over \delta E}\sum\limits_{|E_n-E|<\delta E}\langle E_n|O(u) O(0)|E_n\rangle &=&
    \rho(E)e^{uE}\langle E|O e^{-u H}O|E\rangle\nonumber\\
   \langle E|O e^{-u H}O|E\rangle &=&\int_0^{\infty} ds \rho(s)e^{-{s^2\over 2}u}\scaleto{{|\Gamma(\Delta+i(s+\sqrt{2E}))|^2|\Gamma(\Delta+i(s-\sqrt{2E}))|^2\over 2^{2\Delta+1} \Gamma(2\Delta)}}{25pt}.~~~~~~~~~\label{TwoPFMicro}
\end{eqnarray}
Notice that $|E\rangle$ is not $|E\rangle_G$, the former represents a one side microstate, while the later is a gravitational state.  Accordingly $ \langle E|O(u)O(0)|E\rangle$ stands for a two point function in a microstate.
 To study ETH, we will consider the case of a heavy black hole $E={S^2\over 2}={2\pi^2 \over\beta^2}\gg 1$. From the discussion in last section, we know that the matrix element tries to concentrate $s$ around $\sqrt{2E}$ and thus we can approximate $\rho(E)\rho(s)|\Gamma(\Delta+i(s+S))|^2$ as proportional to $sS^{2\Delta-1}e^{\pi(s+S)}$. Using integral representation for $|\Gamma(\Delta+i(s-S))|^2 $ we derive the two point function in microcanonical essemble with energy $E$ is proportional to:
\begin{equation}
{\rho(E)S^{2\Delta-1}\over u^{3/2}}\int d\xi (\pi+i\xi)e^{-{2\over u}(\xi-i{\pi\over 2})^2-(\pi+2i\xi) S+u{S^2\over 2}}{1\over(\cosh \xi)^{2\Delta}}.
\end{equation}
 The $\xi$ variable can be understood as the measure of time separation in units of effective temperature between two operators and its fluctuation represents the fluctuation of the effective temperature.  
 And the final integral can be understood as a statistical average of correlation functions with different temperatures.  
 If we put back the Newton Constant $G_N={1\over N}$, we have $S\sim N$ and $u\sim N^{-1}$ (\ref{BCond}).
 As can be seen from the probability distribution, the fluctuation is of order ${1\over \sqrt{N}}$, and hence for large $N$ system we can use saddle point approximation:
 \begin{equation}
    \xi=i({\pi\over 2}-{S\over 2}u) -{u\Delta \tanh \xi\over 2}=i({\pi\over 2}-{\pi u\over \beta})-{u\Delta\tanh \xi\over 2}.
 \end{equation}
 The first piece gives the typical temperature of the external state,
 while the last piece comes from the backreaction of operator on the geometry.  If we first take the limit of large N, one simply get that the two point function in microcanonical essemble is the same as canonical essemble.  
 However, the euclidean correlator in canonical essemble is divergent as euclidean time approach to inverse temperature $\beta$ because of KMS condition.  
 Such singular behavior plays no role in the microcanonical essemble so is called a ``forbidden singularity" in ETH \cite{Fitzpatrick:2016ive, Faulkner:2017hll}. In our situation we can see directly how the forbidden singularity disappears in the microcanonical essemble. 
 When $\xi$ approach $-i{\pi\over 2}$ at the forbidden singularity the backreaction on the geometry becomes large and hence the effective temperature becomes lower:
 \begin{equation}
 {2\pi\over\beta*}\rightarrow {2\pi\over\beta}-{\Delta \over \pi-{\pi u\over \beta}}.
 \end{equation}
 At the time ${\beta-u\over\beta}\sim{\Delta\over N}$, the backreaction is important and we expect to see deviation from thermal correlators.
Therefore the correlation function in microcanonical essemble will never have singularity away from coincide point.
\subsection{Three Point Function}
\begin{figure}
    \centering
    \includegraphics[width=0.5\textwidth]{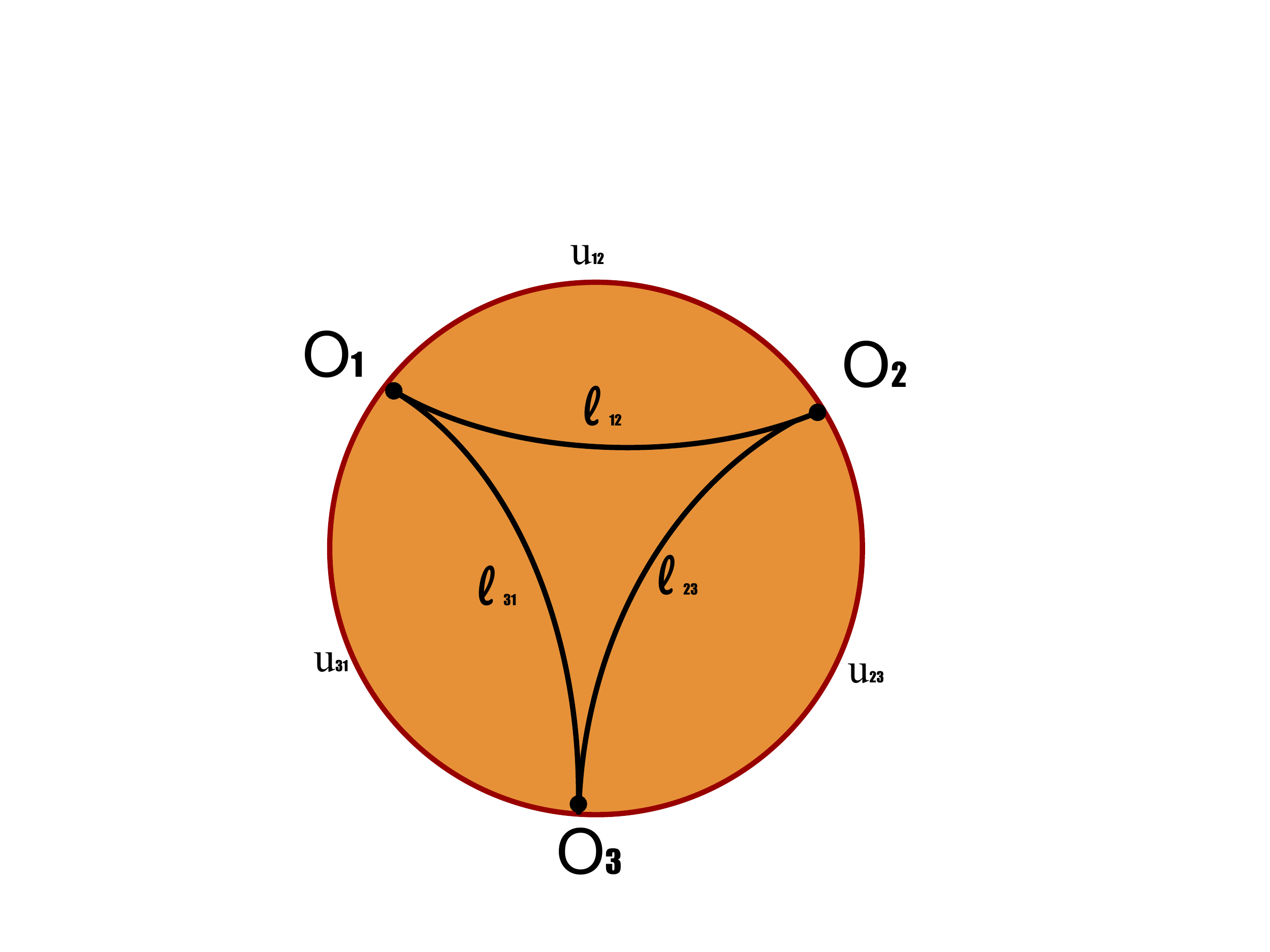}
    \caption{Bulk Diagram for Three Point Function}
    \label{fig:threepointfunction}
\end{figure}

The bulk diagram of the three point function will be like Figure \ref{fig:threewavefunction} with additional operator inserting at the intersection points (See Figure \ref{fig:threepointfunction}).
The QFT three point correlation function in $AdS_2$ is fixed by conformal symmetry and we can write it down as:
\begin{equation}
\langle O_1 O_2 O_3\rangle=C_{123}{z_1^{\Delta_1} z_2^{\Delta_2} z_3^{\Delta_3}\over |x_{12}|^{\Delta_1+\Delta_2-\Delta_3}|x_{23}|^{\Delta_2+\Delta_3-\Delta_1}|x_{13}|^{\Delta_1+\Delta_3-\Delta_2}}={C_{123}\over \ell_{12}^{\Delta_1+\Delta_2-\Delta_3}\ell_{23}^{\Delta_2+\Delta_3-\Delta_1}\ell_{13}^{\Delta_1+\Delta_3-\Delta_2}}.
\end{equation}
$\Delta_i$ is the conformal dimension of $O_i$. Putting them in formula \ref{final formula} and rewrite the propagator in terms of the wavefunction (\ref{wavefunction}), we have the quantum gravitational three point function:
\begin{equation}
\langle O_1 O_2 O_3\rangle_{QG}=\int_{x_1>x_2>x_3}{\prod_{i=1}^3{dx_i dz_i}\over \text{V(SL(2,R))}}\Psi_{u_{12},\ell_{12}}\Psi_{u_{23},\ell_{23}}\Psi_{u_{13},\ell_{13}}I_{\ell_{12},\ell_{23},\ell_{13}}{C_{123}\over \ell_{12}^{\Delta_1+\Delta_2-\Delta_3}\ell_{23}^{\Delta_2+\Delta_3-\Delta_1}\ell_{13}^{\Delta_1+\Delta_3-\Delta_2}}.
\end{equation}
We can view this expression as an inner product of three universe wavefunction with the interior, inserting three bilocal operators $\tilde O_{ij;k}\tilde O_{ij;k}$ with dimension $\tilde\Delta_{ij;k}={1\over 2}(\Delta_i+\Delta_j-\Delta_k)$ between them. One can fix the $SL(2,R)$ symmetry and express the integral in terms of $\ell_{ij}$, it is the same exercise as in open string calculation to find the Jacobian factor (Appendix \ref{gauge fix}).  Here we can just argue that in order to get the partition function at $\Delta=0$, the measure has to be flat.  Therefore the three point function factorizes into form:
\begin{equation}
\small{\langle O_1(u_{1})O_2(u_{2})O_3(u_{3})\rangle_{\text{\rm QG}}}\propto C_{123}\int_0^{\infty}d\tau \rho(\tau)\mathcal{I}_{\tau}(u_{12},\tilde\Delta_{12;3})\mathcal{I}_{\tau}(u_{23},\tilde\Delta_{23;1})\mathcal{I}_{\tau}(u_{31},\tilde\Delta_{31;2})
\end{equation}
while $\mathcal{I}_{\tau}(u_{ij},\tilde\Delta_{ij;k})$ is an integral of $\ell_{ij}$ which gives the two point function in microstate $E_{\tau}$ (\ref{TwoPFMicro}) with the $e^{uE_{\tau}}$ factor stripped off:
\begin{equation}
\mathcal{I}_{\tau}(u_{ij},\tilde\Delta_{ij;k})={1\over 2}\int_{0}^{\infty} d\ell_{ij}\Psi_{u_{ij},\ell_{ij}}{1\over \ell_{ij}^{\Delta_i+\Delta_j-\Delta_k}}K_{2i\tau}({4\over \ell_{ij}})=\langle E_{\tau} |\tilde O_{ij;k}e^{-u_{ij}H}\tilde O_{ij;k}|E_{\tau}\rangle;~~~~~E_{\tau}={\tau^2\over 2}.
\end{equation}
Again the normalization constant can be fixed by choosing $O_i$ to be identity.

\subsection{Einstein-Rosen Bridge}
The Einstein-Rosen Bridge in a classical wormhole keeps growing linearly with time and this behavior was conjectured to related with the growth of computational complexity of the dual quantum state \cite{Stanford:2014jda}.
 Based on the universal behavior of complexity growth, Susskind proposed a gravitational conjecture in a recent paper \cite{Susskind:2018fmx} about the limitation of classical general relativity description of black hole interior.  The conjecture was stated as follows:
 
 \textsl{Classical general relativity governs the behavior of an ERB for as long as possible.}

In this section, we will test this conjecture using the exact quantum wavefunction of JT gravity (\ref{wavefunction}).  We will in particular focusing on the behavior of ERB at time bigger than $1$. The size of Einstein-Rosen Bridge $\mathcal{V}$ in two dimensions is the geodesic distance $d$ between two boundaries, and can be calculated in thermofield double state $|u\rangle$ as:
\begin{equation}
    \mathcal{V}=\langle u|d|u\rangle
\end{equation}
We want to focus on the dependence of volume on Lorentzian time evolution. Therefore we do analytic continuation of $u$ in Lorentzian time: $u={\beta\over 2}+it$.  Using the WdW wavefunction in $\ell$ basis (\ref{wavefunction}) and the relation between $d$ and $\ell$, we can calculate the expectation value exactly.  This can be done by taking the derivative of the two point function (\ref{2point function}) with respect to $\Delta$ at $\Delta=0$. Using the integral representation for $|\Gamma(\Delta+i(s_1-s_2))|^2$, the only time dependence of volume is given by: 
\begin{equation}
    \mathcal{V}(t)={1\over \mathcal{N}}\int_{-\infty}^{\infty}d\xi\int_0^{\infty} ds_1ds_2\rho_1\rho_2 e^{i(s_1-s_2)\xi-i(s_1-s_2){(s_1+s_2) \over 2}t-{\beta\over 4}(s_1^2+s_2^2)}\log(2\cosh{\xi\over 2})|\Gamma(is_1+is_2)|^2.
\end{equation}
The limit we are interested in is $\beta\ll 1\ll t$ \footnote{Remember that we are measuring time in units of $\phi_r$ (\ref{BCond}), so time order $1$ is a quantum gravity region.}, in which case the integral has a saddle point at \footnote{Actually this saddle point is valid for any range of $t$ as long as $\beta\ll 1$.}:
\begin{equation}\label{SaddlePoint}
s_{1,2}={2\pi\over \beta};~~~~~~~~~\xi={2\pi t\over \beta}.
\end{equation}
Therefore the volume has linear dependence in time:
\begin{equation}
\mathcal{V}(t)\sim{2\pi t\over \beta}.
\end{equation}
Using the complexity equal to volume conjecture \cite{Stanford:2014jda,Brown:2015bva}, the complexity of thermofield double state is proportional to the maximum volume:
\begin{equation}
\mathcal{C}(t)=\#\mathcal{V}(t)=\#{2\pi t \over\beta} .
\end{equation}
The proportionality constant is suggested in \cite{Brown:2018kvn} to be $S_0$ based on classical calculation of near extremal black hole. This, however, is not very clear in our model since $S_0$ is the coupling constant of the pure Einstein-Hilbert action and decouples with JT theory (\ref{full action}).
Since the saddle point (\ref{SaddlePoint}) is actually valid from early time to late time, the proportionality constant can be fixed at classical level and once we fix it we can conclude that \textsf{the length of Einstein-Rosen Bridge (or complexity of the state) keeps linearly growing even considering quantum gravity effects in JT theory.} 
We want to comment that this is not an obvious result that one can expect from classical observables. 
For example, one might argue that we can extract the information of the ERB from two sided correlators for the reason that semiclassically we can approximate the correlator as $e^{-m d}$. 
Therefore one can conclude the ERB has linear growth from the quasinormal behavior of the correlator.
However such observables can only give us information of ERB up to time order $1$, which is the same time scale we can trust the classical general relativity calculation.  
After that the correlation function changes from exponential decay into universal power law decay ${1\over t^{3}}$ as one can directly derive from analytic continuation of result (\ref{2point function}).  
If we still use such correlator to extract information about ERB we would get the wrong conclusion that it stops its linear growth after time order $1$.
The reason why it is incorrect is that at this time scale the operator disturbs the state and causes different energy states interfere each other strongly.  It is simply that the correlator can no longer be described by the classical geometry, rather than the interior stops to behave classically.
From our calculation, we see that if we probe of the state in a weaker and weaker way, we are still able to see the classical geometry.
Lastly, we want to talk a little about when JT gravity needs to be modified.  A naive estimate can be made from the partition function that when $\beta$ approaches $e^{{2\over 3}S_0}$, the partition function becomes less than one and definitely at this time scale we need new physics. A recently study of gravitational physics at this time scale was discussed in \cite{Saad:2018bqo}.

\section{Conclusion}

Our result gives an explicit formula (\ref{final formula}) to calculate all order corrections to correlation functions from quantum gravity in two dimensions.
The formula can be understood diagrammatically and we call it Gravitational Feynman Diagram.
We also give the exact Wheeler-DeWitt wavefunction and discuss the growth of its complexity quantum mechanically.

Although we are focusing on theoretical description of two dimensional black holes, the near-extremal black holes in nature should contain these features.
Both Reissner–Nordström black holes and Kerr black holes have an $AdS_2$ throat near their extremality.
For those black holes, the gravitational effects are enhanced by the their near extremal entropies (the coupling constant is $\phi_h$ rather than $\phi_0+\phi_h$) and therefore are better backgrounds to test gravitational effects.
We should however point out that the observational black holes all have large near extremal entropies and thus are very classical \cite{Preskill:1991tb}.
In addition, the Thorne limit of Kerr black hole sets a lower bound on the near extremal entropy in nature.
But for the Primordial black holes in early universe, our story might play a role and it will be interesting to study the physical consequence in that situation.

Another application of our result is to connect with SYK type models \cite{Maldacena:2016hyu,Kitaev:2017awl,Witten:2016iux,Klebanov:2016xxf} since those models all have an emergent Schwarzian action at low energy.  On that account, the exact Schwarzian quantization can be used to test ${1\over N}$ corrections to those models.  For example, one can try to directly test the two point function with SYK numerics \cite{BKobrin} or one can use SYK models to understand the microscopic description of WdW wavefunction and its complexity.  It might also be interesting to consider the black hole information paradox \cite{Fiola:1994ir} and late time traversable wormholes including the quantum fluctuations of the boundary \cite{Gao:2016bin,Maldacena:2017axo,Susskind:2017nto,Yoshida:2017non, Maldacena:2018lmt}.

{\bf Acknowledgments } 

The author wants to give special thanks to J.Maldacena and D.Stanford for patient guidance and help all through the project.
We also want to thank A.Almheiri, L.Iliesiu, J.Jiang, J.Ripley, A.Kitaev, B.Kobrin, R.Mahajan, A.M.Polyakov, S.J.Suh, G.Turiaci and  H.Verlinde for discussions. 
Z.Y is supported by Charlotte Elizabeth Procter Fellowship from Princeton University.

\appendix
\section{Gauge Fix SL(2,R)}
\label{gauge fix}
This section reviews the procedure to fix SL(2,R) gauge which is needed for calculating correlation functions in quantum gravity using formula (\ref{final formula}).
With the parametrization of group elements in SL(2,R) by $g=e^{i\epsilon_{\alpha}L_{\alpha}}$ ($\alpha=\pm 1,0$) near the identity, we have $g$ acting on $\boldsymbol{x}$ as following (\ref{SLTwoLq}):\
\begin{equation}
g x=x-\epsilon_{-1}-\epsilon_0 x+\epsilon_1 x^2;~~~~~~g z=z-\epsilon_0 z+\epsilon_1 2xz.
\end{equation}
Choosing the gauge fixing condition as $f_{\alpha}(g \boldsymbol{x})=0$, we can fix the SL(2,R) symmetry in (\ref{final formula}) using Faddeev-Popov method.
 First we have the identity:
 \begin{equation}
 1=M(\boldsymbol{x})\int d g \delta(f_{\alpha}(g \boldsymbol{x}))
 \end{equation}
Because the measure is invariant under group multiplication, $M(\boldsymbol{x})$ is equal to $M(g\boldsymbol{x})$ and we can calculate it at the solution $\boldsymbol{x}_0$ of the gauge constraints on its orbit: $f_{\alpha}(\boldsymbol{x}_0)=0$, $\boldsymbol{x}_0\in G(\boldsymbol{x})$.  Since the Haar measure is flat near the identity we have \footnote{The normalization constant is arbitrary and we choose it to be one.}:
\begin{equation}
M(\boldsymbol{x})=\det\left({\partial f_{\alpha}(g\boldsymbol{x}_0)\over \partial \epsilon_{\beta}}\right)\bigg\rvert_{\epsilon_{\beta}=0}.
\end{equation}
Inserting $1$ in integrals of SL(2,R) invariant function $F(\boldsymbol{x})$ like the one in (\ref{final formula}), we have:
\begin{equation}
\int d\boldsymbol{x} F(\boldsymbol{x})=\int d\boldsymbol{x}M(\boldsymbol{x})\int dg\delta(f_{\alpha}(g \boldsymbol{x}))F(\boldsymbol{x})=\int dg \int d\boldsymbol{x}M(\boldsymbol{x})\delta(f_{\alpha}(\boldsymbol{x}))F(\boldsymbol{x}).
\end{equation}
We see that the volume of SL(2,R) factorizes out and we have the gauge fixed expression:
\begin{equation}
\int d\boldsymbol{x}\det\left({\partial f_{\alpha}(g\boldsymbol{x}_0)\over \partial \epsilon_{\beta}}\right)\bigg\rvert_{\epsilon_{\beta}=0}\delta(f_{\alpha}(\boldsymbol{x}))F(\boldsymbol{x}).
\end{equation}

\section{Wavefunction with matter}
\begin{figure}
    \centering
    \includegraphics[width=0.4\textwidth]{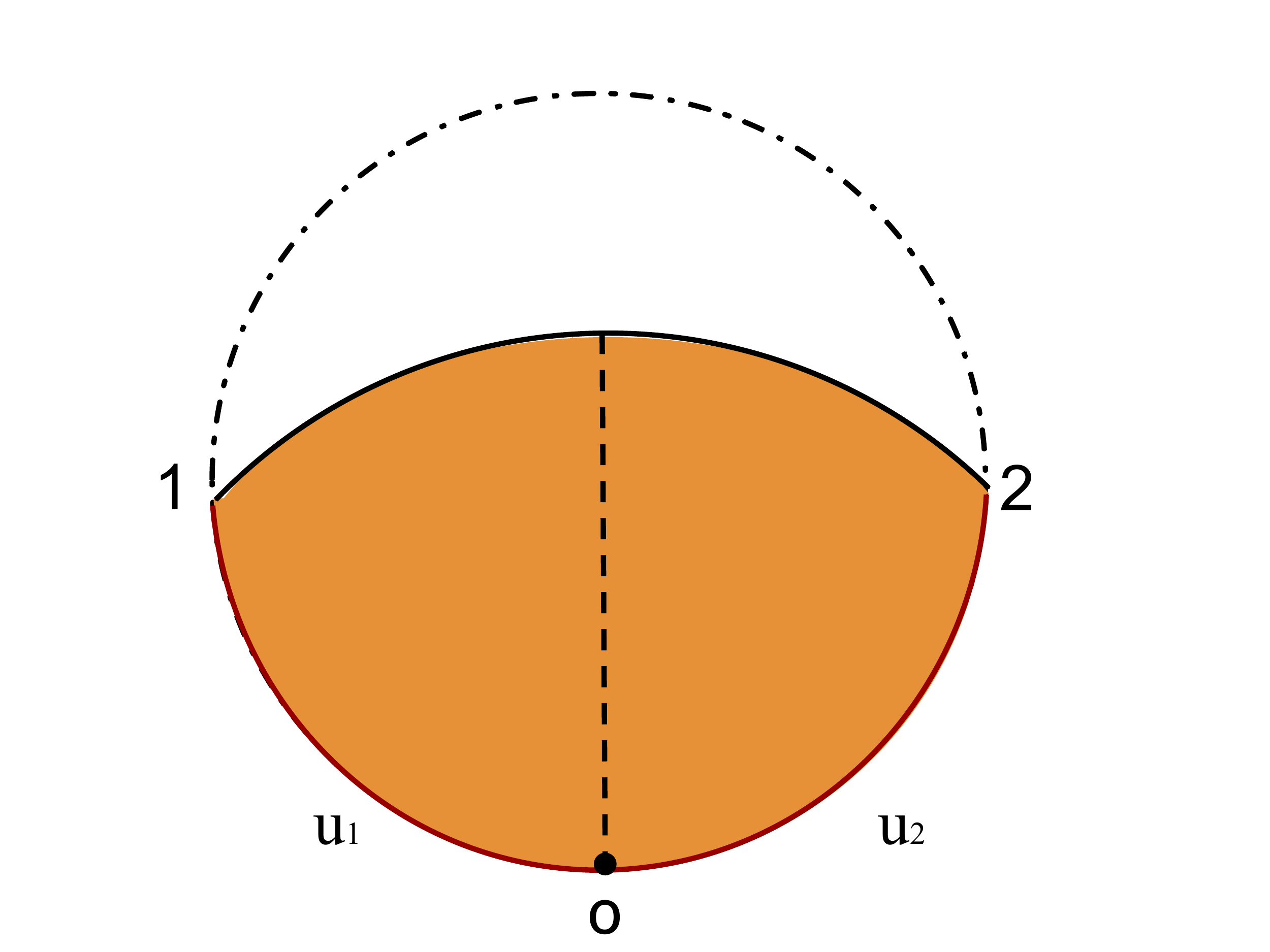}
    \caption{WdW wavefunction with matter}
    \label{fig:WdWMatter}
\end{figure}

Including matter sector in JT gravity (action \ref{full action}), we can discuss the exact wavefunction including matter backreaction.  Schematically, since the geometry on which the matter field propagates is not changed, the WdW wavefunction $\Phi$ including matter sector will be :
\begin{equation}
|\Phi\rangle=\sum_{n}|\Psi_n\rangle \otimes|n\rangle,
\end{equation}
where $|n\rangle$ denotes the matter state in fixed AdS background, and $|\Psi_n\rangle$ means the gravitational wavefunction after backreaction from matter state $|n\rangle$.
By specifying the boundary condition of the matter in Euclidean evolution, one can create different types of states such as vacuum state. 
The vacuum state in AdS is stable and will not backreact on the gravity sector and thus one will get the same story discussed in section \ref{WdW}. 
We will consider another type of state that is created by inserting one boundary operator $O$ during the euclidean evolution (Figure \ref{fig:WdWMatter}).
The operator $O$ creates a single SL(2,R) representation with conformal dimension $\Delta$. 
For the reason that the boundary is fluctuating, $O$ does not create only one asymptotic state, but a superposition of its descendants: $|\Delta, n\rangle$.  The transition amplitude from $|O(\boldsymbol{x})\rangle$ to $|\Delta,n\rangle$ can be determined from the asymptotic behavior of two point function:
\begin{equation}
\lim_{x'\rightarrow \infty}\sum_n\langle O(x')|n\rangle\langle n|O(x)\rangle={z'^{\Delta}z^{\Delta}\over |x-x'|^{2\Delta}}={z'^{\Delta}z^{\Delta}\over x'^{2\Delta}}(1+2\Delta {x\over x'}+...+{\Gamma(2\Delta+n)\over \Gamma(2\Delta)\Gamma(n)}{x^n\over x'^n}+...)
\end{equation}
Therefore we have:
\begin{equation}
|O(x)\rangle=\sum_{n}\sqrt{\Gamma(2\Delta+n)\over \Gamma(2\Delta)\Gamma(n)}z^{\Delta}x^{n}|\Delta,n\rangle
\end{equation}
Notice that because the matter carries SL(2,R) charge, the gravitational part is not a singlet and in particular will depend on the location of two boundary points $\boldsymbol{x_1}$ and $\boldsymbol{x_2}$. By choosing our time slice to be the one with zero extrinsic curvature, we can get the backreacted gravitational wavefunction:
\begin{equation}
\Psi_n(\boldsymbol{x_1},\boldsymbol{x_2})=e^{-2{z_2+z_2\over x_2-x_1}}\int d\boldsymbol{x}\tilde K(u_1,\boldsymbol{x_1},\boldsymbol{x})\tilde K(u_2,\boldsymbol{x},\boldsymbol{x_2})\sqrt{\Gamma(2\Delta+n)\over \Gamma(2\Delta)\Gamma(n)}z^{\Delta}x^{n}.
\end{equation}

\section{Boundary effective action from CFT}
\label{appendix: CFT effective action}
CFT partition function in two dimension has simple dependence on the shape of geometry by Liouville action.  More precisely, the CFT partition function of central charge $c$ on geometries related by $g=e^{2\rho}\hat g$ is related:
\begin{equation}\label{Liouville}
  Z[g]=e^{{c\over 6}S_L}Z[\hat g];~~~~~~~~~~~~  S_L={1\over 4\pi}\Big[\int (\hat\nabla \rho)^2+\rho\hat R +2\int \rho \hat K \Big].
\end{equation}
Our strategy to get the effective action of the boundary shape is to first find a conformal map that maps the boundary into a circle, and then evaluate the Liouville action on that new metric. 
By Cauchy's Theorem, such a conformal map always exists and is uniquely determined up to SL(2,R) transformation.
The SL(2,R) transformation does not change the weyl factor and therefore does not affect our final result.
The original metric has constant negative curvature and we will parametrize it by a complex coordinate $h$ as ${4\over (1-|h|^2)^2}d h d\bar h$. If we denote the conformal map as $h(z)$, where $z$ is the coordinate in which the boundary is a circle $|z|=1$, then the new metric in coordinate $z$ is:
\begin{equation}
ds^2={4\partial h \bar\partial \bar h\over (1-|h|^2)^2}dzd\bar z.
\end{equation}
The holomorphic function $h(z)$ determines the boundary location in $h$ coordinate (parametrized by $u$) at \footnote{$z= e^{i\theta}$ and $h=r e^{i \tilde \theta}$}:
\begin{equation}
r(u)e^{i\tilde\theta (u)}=h(e^{i \theta(u)}),
\end{equation}
where $r(u)$ and $\tilde\theta(u)$ are related by the metric boundary condition:
\begin{equation}
{4(r'^2+r^2\tilde\theta'^2)\over (1-r^2)^2}=q^2;~~~~~~r(u)=1-q^{-1} \tilde\theta'(u)+O(q^{-2}).
\end{equation}
Combine these two equations at large $q$ we get a Riemann Hilbert type problem:
\begin{equation}
e^{i\tilde\theta(u)}(1-q^{-1} \tilde\theta'(u))=h(e^{i\theta(u)}).
\end{equation}
This equation can be solved by the holomorphic property of $h(z)$ and the solution is:
\begin{equation}
h(z)=z\left(1-{1\over 2\pi q}\int d\alpha {e^{i\alpha}+z\over e^{i\alpha}-z}\theta'(\alpha)\right)
\end{equation}
Choosing our reference metric $\hat g$ to be flat, we have:
 \begin{equation}
     \rho={1\over 2}(\log{\partial h}+\log\bar\partial\bar h)-\log(1-h\bar h).
 \end{equation}
Evaluation of the Liouville action (\ref{Liouville}) is then straightforward and gives us a Schwarzian action:
\begin{equation}
S_L=-{1\over 2}-{1\over 4\pi q}\int du Sch(\tan({\theta\over 2})).
\end{equation}
We want to remark that the sign in front of the Schwarzian action is negative so a naive attempt to get induced gravity from large number of quantum fields does not work.\footnote{For an other interpretation of this result, see \cite{Callebaut:2018nlq}. Matter quantization with JT theory was also considered in \cite{CHAMSEDDINE199298} in the context of non-critical string theory.}

\section{Connection with the relativistic particle and Pair Production}
\label{appendix:relativistic particle}
We will start from a formal expression for the relativistic particle with mass $m$ and charge $q$ in an electric field and 
gradually implement these changes to get the partition function of the JT theory. 
 The partition function for the relativistic particle has the form 
\be \la{Zrel}
Z_{\rm rel}(m_0,q) = \sum_{\rm Paths}  e^{ -m_0 L } e^{ - q \int a } = \int_0^\infty  {d \tau \over \tau } e^{ - \half \tau \mu^2 } \int {\cal D} x {\cal D} y    \exp\left(  - \int_0^\tau d \tau' \half { \dot x^2 + \dot y^2 \over y^2 } - q \int {d x \over y } \right)
\ee
where $L$ is the length of the path. In the right hand side 
$\tau$ is Schwinger's proper time, which is related by a  renormalization factor to the actual proper time of 
the path \cite{Polyakov:1987ez} (Chapter 9). Also, we have that 
$\mu^2 = {( m_0 -m_{cr}) \over \tilde \epsilon } $, where $\tilde \epsilon$ is a UV cutoff for the path integral (not to be confused with $\epsilon$ in \nref{BCond}). 
If we are interested in the JT partition function at finite temperature, then we are interested in fixing the length
of the paths. As we mentioned, this is the same as fixing the Schwinger time in \nref{Zrel}. 
More explicitly, we can multiply $Z_{\rm rel}(m_0,q)$ by $e^{ m_0  \beta \over \tilde \epsilon }$ and then integrate
over $m_0$ along the imaginary axis (with a suitable) real part to fix the length of the path. 
This then gives $ \beta = \tau $ in the above expression. 
The precise value of $\mu^2$ can be absorbed by shifting the ground state energy. 
It will be convenient for further purposes to set $\mu^2 = q^2 - 1/4$.  
 The path integral in the right hand side of \nref{Zrel} has an infinite
volume factor. We drop this factor when we divide by the volume of $SL(2)$. In addition, the factor of $1/\tau$ should be dropped
because we view configurations that differ by a shift in proper time as inequivalent. 
After all these modifications we find 
\be \la{JTPa}
Z_{JT}(\beta)  = e^{S_0} e^{  2 \pi q }  {1\over 2\pi} G(\beta , x,y; x,y)
\ee
where $G$ denotes the propagator of the non-relativistic problem of a particle in an electric 
field 
\be
G(\tau ; x,y; x' ,y') = \langle x,y| e^{ - \tau H } |x', y' \rangle = 
\int { \cal D} x { \cal D } y \exp\left(  - \int_0^\tau d \tau' \half { \dot x^2 + \dot y^2 \over y^2 } - q \int {d x \over y } \right)
\ee

At first sight, the statement that a gravitational system is equivalent to a particle makes no sense, since we know that the entropy of a particle is very small. Usually the partition function is of the form $Z|{\text particle}\sim (\hbar)^{\#}$ for a particle system, but black hole has entropy of order ${1\over \hbar}$, that is $Z|_{BH}\sim e^{\#\over \hbar}$.  This is because in the particle case, the major contribution in functional integral is given by stationary solution, and the fluctuations near the stationary solution give the power of $\hbar $, while for the gravitational system, a stationary solution will corresponding to no geometry and we have the requirement of the boundary should have winding number one.  A solution with winding number one in the particle system is an instanton contribution for particle pair production, which is usually very small and is in addition imaginary, so how can this matches with gravity system? The pair creation rate for a particle with charge $q$ and mass $m$ in AdS can be estimated from Euclidean solution which is a big circle with radius $\rho_b=\text{arctanh}({m\over q})$: 
\begin{equation}\label{instanton effect}
    I=mL-qA\sim 2\pi m \sinh \rho_b -2\pi q(\cosh \rho_b -1)+\pi(m \sinh\rho_b-q\cosh\rho_b)\delta\rho^2\sim  2\pi q.
\end{equation}
We see that the damping factor is exactly cancelled out by our gravitational topological piece.  The negative norm mode is related with the rescaling of the circle. That is not allowed in canonical essemble because of the temperature constraint.

\section{Details on the Schwarzian limit of Propagator}
\label{appendix:large q}
The main technical difficulty in finding the large $q$ limit of propagator (\ref{particle heat kernel}) is the hypergeometric function. To properly treat it, we can first use transformation of variables:
\begin{eqnarray}\label{hypergeometric function simplify}
     {1\over d^{1+2is}}F({1\over 2}-iq+is,{1\over 2}+iq+is,1,1-{1\over d^2})=~~~~~~~~~~~~~~~~~~~~~~~~~~~~~~~~~~~~~~~~~~~~~~~~~~~~~~~~\nonumber\\
     {\Gamma(-2is)\over d^{1+2is} \Gamma({1\over2}-is+iq)\Gamma({1\over 2}-is-iq)}F({1\over 2}+is-iq,{1\over 2}+is+iq,1+2is,{1\over d^2})+(s\rightarrow -s)~~~~~
\end{eqnarray}
   In the limit of large $q$ ($d$ scales with $q$), we have approximation of hypergeometric function:
   \begin{equation}
       F({1\over 2}+is-iq,{1\over 2}+is+iq,1+2is,{1\over d^2})\sim \Gamma(1+2i s)({d\over q})^{2i s}I_{2i s}({2 q\over d}).
   \end{equation}
Using reflection property of gamma function together with large $q$ approximation of $\Gamma({1\over 2}-is+iq)\Gamma({1\over 2}-is-iq)\sim 2\pi e^{-\pi q}q^{-2i s}$, we have:
\begin{equation}
  (\ref{hypergeometric function simplify})\sim-{ie^{\pi q}\over 2\sinh (2\pi s) d}(I_{-2is}({2q\over d})-I_{2i s}({2q\over d}))={e^{\pi q}\over \pi d}K_{2is}({2q\over d}).
\end{equation}
Putting everything together will give us (\ref{heat kernel}).

\section{Trajectories in Real Magnetic Field}
\label{appendix:landau level}
\begin{figure*}
  \centering
  \subfigure[Landau Level]{%
    \includegraphics[width=0.4\textwidth]{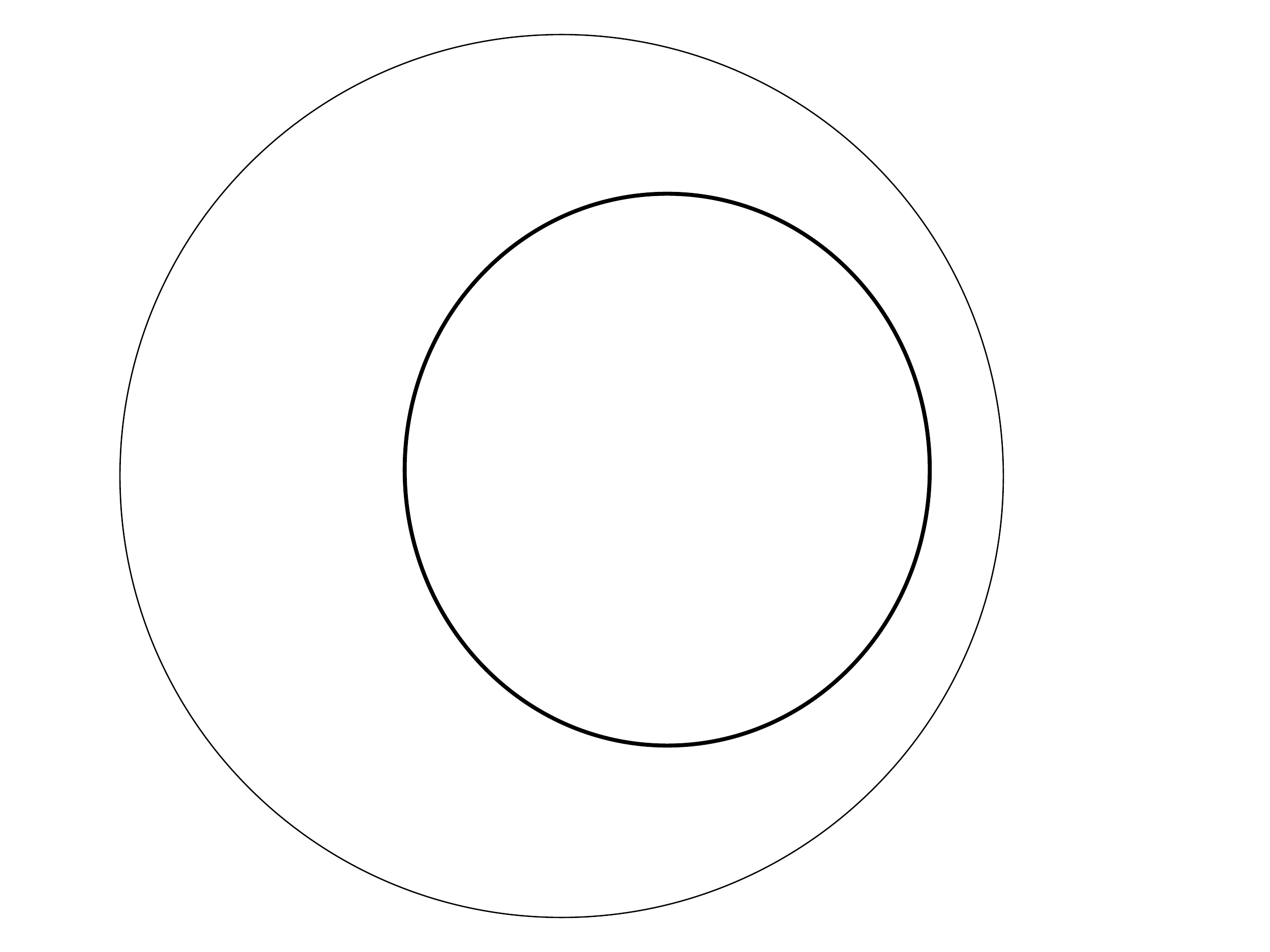}%
    \label{fig:landau level}%
    }\hspace{2cm}
    \subfigure[Scattering State]{%
    \includegraphics[width=0.4\textwidth]{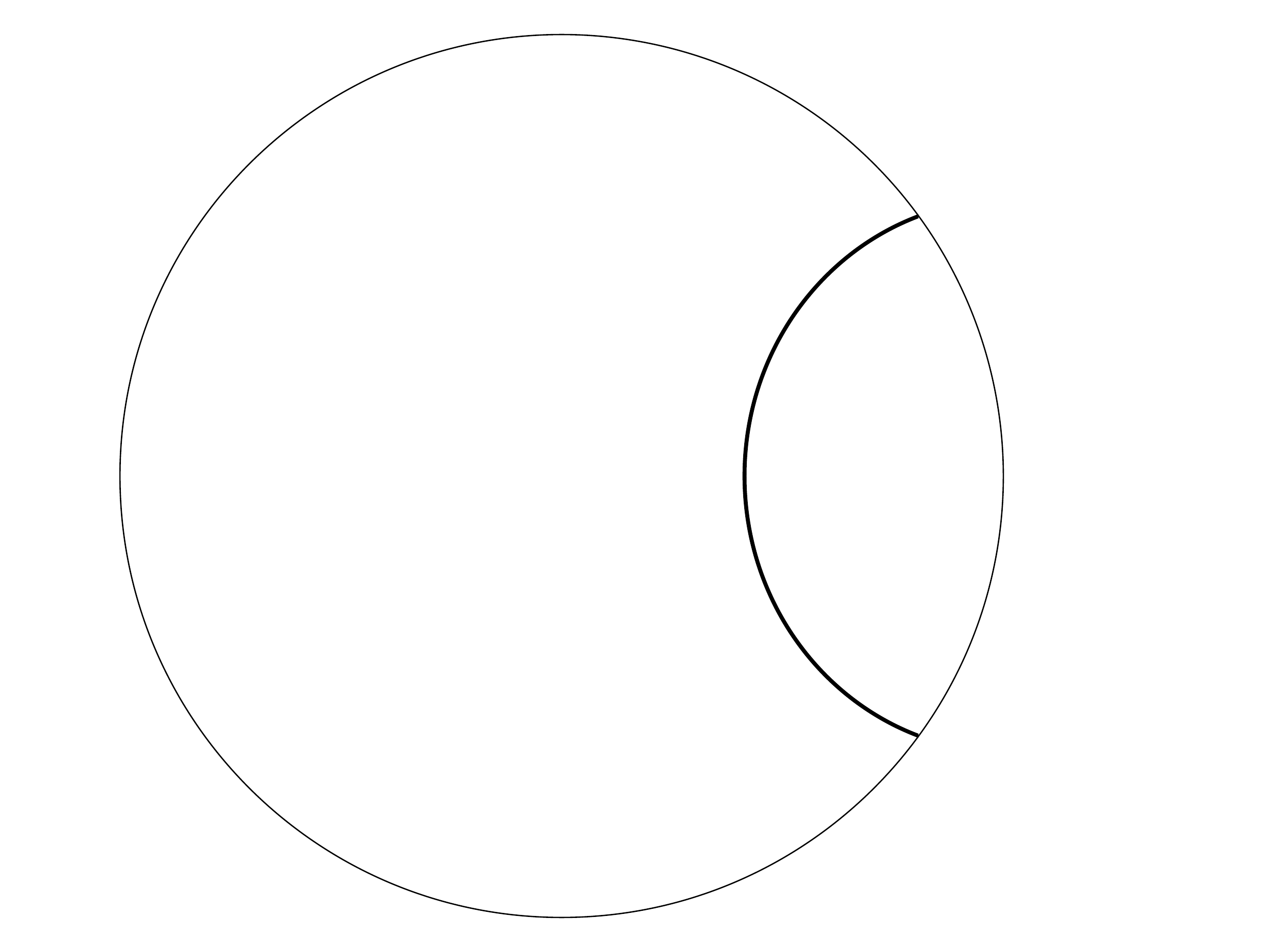}%
    \label{fig:scattering state}%
  }%
  \caption{Particle in real magnetic field} 
  \label{fig:particle in real magnetic field}
\end{figure*}
The equation of motions for a particle in real magnetic field $b$ are:
\begin{eqnarray}
    {\dot x^2+\dot y^2\over 2y^2}=E;\\
    {\dot x\over y^2}+{b\over y}=k,
\end{eqnarray}
where $E$ and $k$ are the conserved energy and momentum respectively.  Since we are only interested in the trajectories, we can introduce a time parametrization $\xi$ such that ${d\tau\over d\xi}={1\over 2k y}$. Then in coordinate $\xi$, we have:
\begin{eqnarray}
     (\partial_{\xi}x)^2+(\partial_{\xi} y)^2={E\over 2k^2};~~~~~~~~
    \partial_{\xi}x=( y -{b\over 2k}).
\end{eqnarray}
This means that we have solutions:
\begin{equation}
    x^2+(y-{b\over k})^2={E\over 2k^2}.
\end{equation}
Those are circles with radius $\sqrt{E\over 2}{1\over k}$ and center at location $(0,{b\over 2k})$. 
So classically we have two types of states as shown in Figure \ref{fig:particle in real magnetic field} : for $E<{b^2\over2}$, the particle is confined by magnetic field and becomes Landau level in the hyperbolic plane; for $E>{b^2\over 2}$, the gravitational effect dominates and particle scatters out of the space.

\bibliographystyle{unsrt}
\nocite{*}
\bibliography{cite}

\end{document}